\documentclass[twocolumn]{article}

\usepackage{arxiv}
\usepackage{geometry}
\usepackage{hyperref}
\usepackage{cite}
\usepackage{enumitem}
\setlist[itemize]{noitemsep, nolistsep}
\usepackage{graphicx}
\graphicspath{{figures/}{pictures/}{images/}{./}} 

\usepackage{amsmath}
\usepackage[ruled,vlined]{algorithm2e}
\usepackage{algpseudocode} 
\usepackage{array}
\usepackage{microtype}
\usepackage{caption}
\usepackage{booktabs}                  
\usepackage[usenames,dvipsnames,table]{xcolor}
\newcommand{\annot}[1]{
\iftrue % uncomment to SHOW paragraph annotations
\textbf{[#1]}
\fi
}
\newcommand{\eg}{e.g.}

\newcommand\rh[1]{\textcolor{black}{#1}}

\newcommand\newText[1]{\textcolor{black}{#1}}
 \newcommand{\qcrFont}[1]{{\fontfamily{qcr}\selectfont{#1}}}
 \newcommand{\comment}[1]{}
 
\title{Nanomatrix:\\ Scalable Construction of Crowded Biological Environments}
\usepackage{authblk}
\author[1]{Ruwayda~Alharbi}
\author[1]{  Ond\v{r}ej~Strnad}
\author[2]{Tobias~Klein}
\author[1]{Ivan~Viola}

\affil[1]{King Abdullah University of Science and Technology (KAUST), Saudi Arabia. E-mails: \{ruwayda.alharbi $\vert$ ondrej.strnad $\vert$ ivan.viola\}@kaust.edu.sa.}

\affil[2]{Nanographics. E-mail: tobias@nanographics.at.}

\date{2022}                   %% if you don't need date to appear
\begin{document}
\twocolumn[
  \begin{@twocolumnfalse}
    \maketitle
    \begin{abstract}
We present a novel method for the interactive construction and rendering of extremely large molecular scenes, capable of representing multiple biological cells in atomistic detail.
Our method is tailored for scenes, which are procedurally constructed, based on a given set of building rules. 
Rendering of large scenes normally requires the entire scene available in-core, or alternatively, it requires out-of-core management to load data into the memory hierarchy as a part of the rendering loop.
Instead of out-of-core memory management, we propose to procedurally \emph{generate} the scene on-demand on the fly. 
The key idea is a positional- and view-dependent procedural scene-construction strategy, where only a fraction of the atomistic scene around the camera is available in the GPU memory at any given time. The atomistic detail is populated into a uniform-space partitioning using a grid that covers the entire scene. Most of the grid cells are not filled with geometry, only those are populated that are potentially seen by the camera. The atomistic detail is populated in a compute shader and its representation is connected with acceleration data structures for hardware ray-tracing of modern GPUs. 
Objects which are far away, where atomistic detail is not perceivable from a given viewpoint, are represented by a triangle mesh mapped with a seamless texture, generated from the rendering of geometry from atomistic detail.
The algorithm consists of two pipelines, the construction-compute pipeline, and the rendering pipeline,
which work together to render molecular scenes at an atomistic resolution far beyond the limit of the GPU memory containing trillions of atoms. We demonstrate our technique on multiple models of SARS-CoV-2 and the red blood cell.

    \end{abstract}
    
  \end{@twocolumnfalse}
]


\section{Introduction}\label{sec:introduction}

\begin{figure*}
     \centering
    \includegraphics[width=1.0\linewidth]{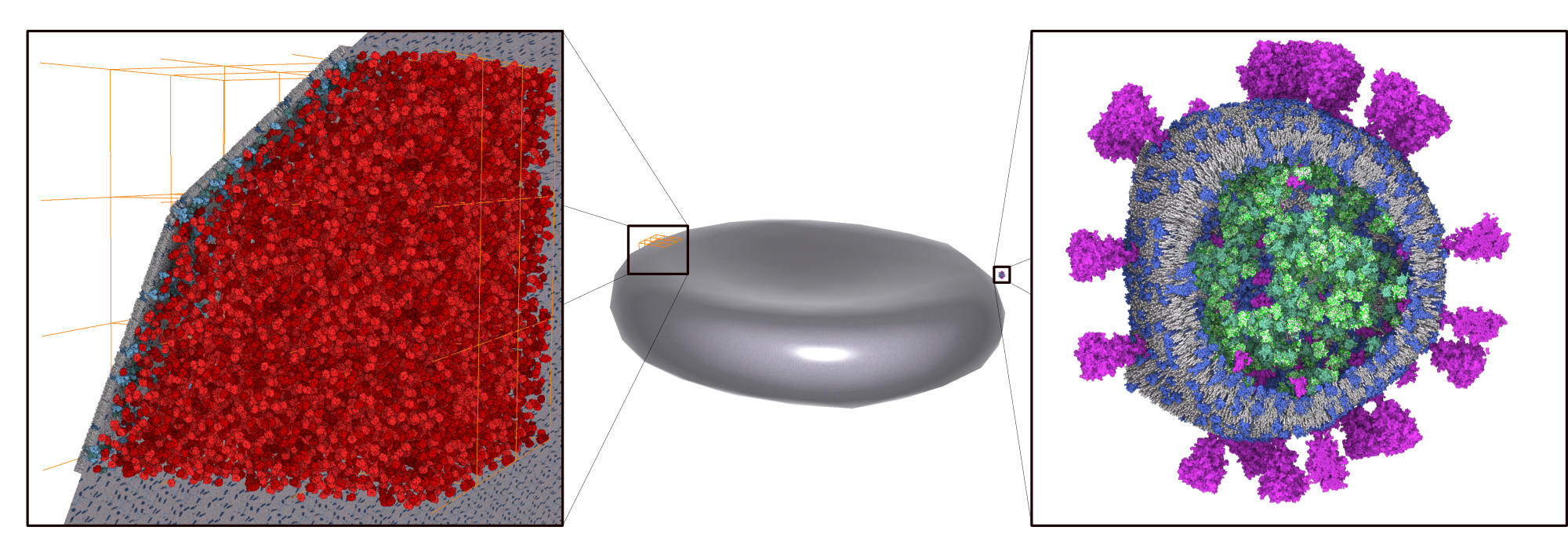}
  \caption{Red Blood Cell model of diameter 8~{\textmu}m containing approx. 250 million of hemoglobin molecules, with lipid-bilayer and membrane-bound proteins (approx. 1.2 trillion atoms in total) constructed and rendered with our view-guided two-level Nanomatrix approach. To the right of the cell there is a model of SARS-CoV-2. The rendering exploits hardware ray tracing, maintaining highly interactive framerates.}
  \label{fig:teaser}
 \end{figure*}

The cellular mesoscale describes the scales that bridge the atomistic nanoscale and the cellular microscale \cite{GOODSELL-2020-Art-and-Science-of-the-Cellular-Mesoscale}. It starts from atoms that form larger molecules like proteins, up to the size where molecules further compose to form viruses, bacteria, or complex multi-compartmental cells. Visualizing biological structures in mesoscale helps biologists to analyze and understand the architecture and functionality of life forms.
%\annot{Why large dataset?}
Mesoscale models of enveloped viruses reach several tens of millions of atoms in size, and these structures are just around 100~nm in diameter. Such a number only describes its macromolecular composition. If we would include water and small molecules, the number of atoms would at least double. Larger models of bacteria or complex cells reach the range of billions or trillions of atoms. \emph{E. coli} for example contains roughly 15 billion C, N, and O atoms (excluding hydrogen) that form its macromolecular composition. Larger structures, such as the red blood cell (RBC) consist of trillions of atoms. Each RBC consists of two-thirds of hemoglobin molecules, which are \rh{circa} 250 million in each RBC. Just the position (3$\times$ \texttt{float}) and rotation (4$\times$ \texttt{float}) information of each hemoglobin instance in a single RBC will amount to 8~GB of data in storage. Storing atomistic information about a complete RBC, including its lipid bilayer membrane, reaches or exceeds the limits of the memory capacity of current consumer GPUs. When attempting to store atomistic-detail information about a larger cell with 50~{\textmu}m in diameter, the required memory will raise to petabytes.
The huge amount of data involved in mesoscale systems poses a challenge for visualization due to hardware limitations. 

Current technologies are able to render biological structures with up to billions of atoms. Our goal is to push this limit and effectively visualize models that contain trillions of atoms. 
Out-of-core approaches usually render massive scenes by keeping a fraction of data in the core memory and loading data from the disc when needed. Instead, we propose a view-guided procedural approach that generates the scene on-demand on the fly. The whole enormous scene will be never completely stored in the memory, only a fraction of it which is close to the viewer. We use a regular grid that uniformly divides the scene into cells. Only cells that are close to the viewer are populated with atomistic/nanoscale geometry, while the remaining cells use an image-based approach to depict the detail, which provides the user with a cellular/microscale description of the far structures. The main contributions include:
\begin{itemize}[noitemsep]
  \item For the nanoscale level, we present new algorithms for the rapid construction of biological structures \newText{which can be directly applied to triangular meshes with arbitrary face sizes.}  
   \item For the cellular level, we propose image-based Wang tiles, derived from geometric Wang tiles, to represent the structures that are far away \newText{ using image to geometry alignment}.
    \item We propose an algorithm for dynamically managing the memory to change between the atomistic and cellular representations.
    \item We propose a parallel rendering scheme that utilizes hardware-accelerated raytracing for molecular visualization.
   % \item  \newText{We propose a view-guided interactive construction approach where the camera trigger GPU to construct a fraction of the scene that is close to the viewer.}
   \item \newText{We demonstrate and evaluate the performance and capabilities of Nanomatrix framework compared to existing visualization methods.}
\end{itemize}
%{In comparison to previous work, these new contributions improve the construction and visualization of biological mesoscale models in several ways. First, our approach is \textbf{general} and \textbf{flexible}. The input for the construction tiles can be generated in multiple ways, depending on the user’s choice of tool. Previous work~\cite{Tobias-2018-Instant-Construction, cellPACK} uses specific packing algorithms that require the scene to be completely constructed before it can be visualized \newText{while our method can easily add/remove elements and change the scene content on-the-fly.} Second, the construction and visualization of our approach are \textbf{scalable} toward larger structures. Building tiles are constructed and placed on demand.} 
\newText{In comparison to previous work, the construction and visualization of our approach are \textbf{scalable} toward larger structures}. Previous work~\cite{Tobias-2018-Instant-Construction, cellPACK} uses specific packing algorithms that require the scene to be completely constructed before it can be visualized \newText{while in our method the scene elements are constructed and placed on demand in realtime}. The visualization follows the same principle and offers geometric detail on demand for closer features and abstraction, in the form of textures, for distant structures. These new features allow users for the first time to interactively generate and visualize biological mesoscale landscapes in the size of a red blood cell in atomistic detail. \newText{Our view-guided construction approach facilitates the interactive exploration of large molecular landscapes. We expect our approach will be useful for communicating scientific discoveries and disseminating scientific knowledge to a broad audience. We see our applicability in a setup like science centers where we take the viewer to an interactive exploration of the molecular universe. Our approach is primarily targeted at biological systems but it can be applicable to any system that exhibits multiscale, multi-instances, dense, 3D nature.}
%=======================
\section{Related Work} 
\label{sec:Related-Work}
The generation of biological mesoscale models is often realized through integrating scientific data from multiple sources into a 'recipe'.
A recipe describes the molecular ingredients and overall composition of the biological environment.
Computer graphics has a long tradition of using structural rules, such as recipes, to generate and visualize man-made and natural phenomena.
A particular problem for very large scenes is that datasets become too massive to fit inside the internal memory.
One solution is to use out-of-core techniques\cite{Varadhan-2002-Out-of-core-rendering-of-massive-geometric-environments,Wald-2019-RTX-Beyond-RayTracing} that store the scene in slow external memory and transmit only currently required data into the fast internal memory.
The input/output communication between the internal and external memory is the bottleneck of such out-of-core approaches.
In this paper, we are using an exclusive in-core approach and trade memory with computing.
Instead of fetching the data from external memory, we construct it on the fly when it is needed.
On-the-fly construction requires very efficient techniques to compute and visualize procedurally generated scenes in real-time.
In this section, we review related work from procedural modeling, texture synthesis, molecular visualization, and parallel rendering.

\paragraph*{\textbf{Procedural Modeling:}} Generating 3D digital content is a very time-consuming and tedious task, while many environments contain self-similar and repetitive structures~\cite{merrell2010model}.
Procedural modeling techniques offer a way to create 3D models from sets of rules or algorithms.
Very early approaches like L-systems use formal grammars to describe natural patterns as they appear in trees, for instance. 
Other phenomena like fire, water, gases, and clouds have also been procedurally generated for decades.
In recent times, computer graphics has experienced significant advances in the procedural modeling of natural as well as man-made structures.
Computer games and movies utilize procedural modeling techniques to create large worlds with varying shapes and styles. 
A fundamental part of such worlds is the automatic modeling of architectures~\cite{wonka2003instant} with different building designs.
Such techniques are capable of generating infinite cities~\cite{greuter2003real} or rich forest scenes~\cite{decaudin2004rendering} in real-time.
In order to generate detailed content interactively, the research focus has shifted towards parallelization~\cite{boechat2016representing,Steinberger2014}, such that computation and memory can be efficiently mapped on graphics hardware.
While most procedural modeling techniques target open worlds, in this work, we focus on a very constrained space where the content is defined through scientific measurements. Procedural modeling in a mesoscale environment is a challenging task. This environment is highly dense with heterogeneous molecular structures in size and shape. There are several modeling techniques that target molecular landscapes, one of which is CellPaint~\cite{Gardner-2018-CellPAINT}. CellPaint allows users to create dynamic molecular illustrations using a 2D style, which shows a cross-section of a biological mesoscale scene. Mesocraft~\cite{Mesocraft} allows the user to interactively specify a set of rules that define the spatial relations of the model's molecules and propagate these rules through the model which results in rapid modeling. {Mesocraft and CellPaint are semi-automatic modeling techniques where the user is expected to participate in the building process of the biological structure, for example, changing the position of a molecule or rotating it. In contrast, our construction approach is fully automatic. Once the user provides the algorithm with the required input, the model is generated without any user interaction.}

Atomistic modeling in mesoscale environments is typically based on packing algorithms, as demonstrated with cellPACK~\cite{cellPACK}.
Packing molecules is a computationally demanding task.
Assembling a 3D mesoscale model from scratch via cellPACK could take from minutes to hours, depending on the size and complexity of the model. 
 Klein et al.~\cite{Klein-2019-Parallel-Generation-and-Visualization-of-Bacterial-Genome-Structures,Tobias-2018-Instant-Construction} propose the {\it instant construction} approach to rapidly create mesoscale models.
It consists of a set of GPU-based population algorithms which generate different types of biological structures.
The approach is limited to structures that fit into GPU memory and focuses on experimenting with different versions and ensembles of mesoscale structures. 
Our \newText{view-guided construction} approach overcomes this limitation, by \newText{populating} structures on the fly once visible. This opens the door to visualizing larger structures like a whole RBC or even infinite molecular landscapes.
\newText{Inspired by Klein et al.~\cite{Tobias-2018-Instant-Construction} work, we use the Wang tiles concept to populate the biological structures. While their tile mapping approach is limited to equally-sized quad-based mesh, we propose a new algorithm that can map tiles into a triangle mesh of varying face sizes. Moreover, we propose a novel algorithm for constructing the inner part of the biological compartment by filling the space with collision-free 3D tiles which eliminates the need for real-time collision handling.}

%Similarly, we present a GPU-based population algorithm; however, our approach differs in several aspects. First, the {\it instant construction} algorithm for populating the membranes required the mesh to be defined as an equally-sized quad-based surface to place the Wang tile on these quads, while our method uses the texture coordinates to map the tile into the triangle mesh of varying size directly. Second, they populate the soluble proteins randomly and then resolve the collisions using a force-based approach while we use the collision-free tile cubes to populate the soluble proteins, which eliminates the need for collision handling and subsequently speed the population process.  

 %+++++++++++++++++++++++++++++++++++++++++++++++++++++++++++++++++++++++++++++++++++++++++++++++++++++++++++

\paragraph*{\textbf{Texture Synthesis:}}
Our approach of generating patches of geometry is very related to the synthesis of textures, especially in the context of Wang tiles. 
Originally, the Wang tiling concept~\cite{wang1961proving} has been proposed as a formal system to cover an infinite plane with non-periodic patterns from a small set of tiles.
The concept is based on tiles with color-encoded edges, whereas each tile is arranged in a way that its edge color matches the adjacent neighbor.
Later adaptions of this approach have been used to map 2D textures onto 3D geometry.
Fu and Leung~\cite{fu2005texture} have extended the concept to apply Wang tiles on arbitrary topological surfaces. Li-Yi~\cite{Li-Yi-2004-Tile-Based-Texture-Mapping-on-Graphics-Hardware} avoid storing large textures on graphics processors by presenting a Wang tiles-based texture mapping algorithm that generates large virtual textures directly on the GPU. Cul{\'i}k and Kari\cite{Culik-1995-An-Aperiodic-Set-of-Wang-Cubes} introduce Wang cubes with color-encoded faces which is a generalization of Wang tiles in 3D. Doškář et al.~\cite{Doskar-2020-Level-set-Based-Design-of-Wang-Tiles-for-Modelling-Complex-Microstructures} use Wang cubes to generate compressed representations of complex microstructural geometries.
Considering that such approaches are constrained to map color information to tiles, in this work we target to map geometry to tiles.
The major challenge of the application of Wang tiles lies in the generation of the tiles, as well as handling the issue of 
%curvature and 
overlapping geometry in the 3D case. 
Comparable approaches with geometry lack performance or use regular patterns~\cite{fleischer1995cellular}. 

%+++++++++++++++++++++++++++++++++++++++++++++++++++++++++++++++++++++++++++++++++++++++++++++++++++++

%+++++++++++++++++++++++++++++++++++++++++++++++++++++++++++++++++++++++++++++++++++++++++++++++++++++++++++++
\paragraph*{\textbf{Large-scale Molecular Visualization:}}
There is a variety of tools for molecular visualization, such as VMD~\cite{VMD}, or PyMOL~\cite{PyMOL}. These tools have been designed for molecules with up to thousands of atoms and are not suited when the data size exceeds tens of millions of atoms~\cite{Knoll-2013-Ray-Tracing-and-Volume-Rendering-Large-Molecular-Data-on-Multi-Core-and-Many-Core-Architectures}. Similarly, generic visualization tools like Amira~\cite{Amira} prioritize generality over scalability and thus reach their limits with increasing dataset sizes. Waltemate et al.~\cite{Waltemate-2014-Membrane-Mapping} map lipid bilayers onto a mesh geometry with interactive rendering performance for up to 10 million atoms on NVIDIA GeForce GTX 770. {Mol* Viewer~\cite{Sehnal21} presents a powerful web-based application for visualizing molecular data. It is able to render an HIV model that contains 67 million atoms smoothly using a NVIDIA RTX 2060 graphics card.} Megamol~\cite{Grottel-2015-MegaMol, OSMegamol} is a visualization framework designed to address interactive visualization of large particle-based datasets. The system is able to render up to 100 million atoms at interactive framerates, which represents in biology scale a virus or a small bacterium. 
%Avizo\cite{} and ParaView

Recently, Ibrahim et al.~\cite{Ibrahim-2021-ProbabilisticParticleOcclusionCulling} introduce a probabilistic occlusion culling architecture using meshlets for acceleration. The algorithm was able to render 232 million particles in 41 frames per second (FPS) on an NVIDIA RTX Titan 24GB.
%\annot{large-scale Molecular Visualization}
Lindow et al.~\cite{Lindow-2012-Interactive-Rendering-Of-Materials} have first presented interactive visualization of large-scale biological data that consists of several billions of atoms. As the biological models often consist of a large number of recurring molecules of a few proteins, they create for each protein a 3D grid structure containing all its atoms and store the grid on the GPU as a 3D voxel and then utilize instancing to repeat the proteins in the scene. In the rendering stage, they draw the instances' voxel and perform raycasting in the fragment shader. Their method was able to render 4,025 microtubules with approximately 10 billion atoms with at least 3 FPS on a NVIDIA GTX 285. 
Falk et al.~\cite{Falk-2013-Atomistic-Visualization-of-Mesoscopic} extend this work by optimizing the depth culling and the rendering method and they add an implicit Level of Detail (LoD) approach. They were able to render 25 billion atoms in 3.6 FPS using a NVIDIA GTX 580. 

Le Muzic et al.~\cite{LeMuzic-2014-Illustrative-Visualization-of-Molecular-Reactions-using-Omniscient-Intelligence}, introduced an optimized approach using a straightforward LoD scheme that does not require grid-based supporting structures. Instead, it relies on the tessellation shader to dynamically inject sphere primitives in the rasterization pipeline for each molecular instance. 
%They were able to render up to 30 billion of atoms at 10 fps.% on an NVIDIA GTX Titan.
Later, they extended their work and introduced cellVIEW~\cite{LeMuzic-2015-cellVIEW} where they reduce the number of injected sphere primitives into the rendering pipeline. % and they were able to render up to 15 billion of atoms at 60 fps. % on an NVIDIA GTX Titan.% NVIDIA GTX Titan X graphics card with 12GB of video RAM. 
The cellVIEW system has set a benchmark with its ability to render 250 copies of an HIV virus model in blood plasma at 60 FPS on NVIDIA GTX Titan. This scene contains 16 billion atoms (each replica contains around 64 million atoms). The goal of our proposed approach is to push this limit and visualize a large biological model that contains trillions of atoms. All of the above-mentioned techniques use procedural impostors for representing atoms to simplify the geometry and accelerate the rendering~\cite{Tarini06,Michel20}. 

\newText{The evolution in computational capabilities has reshaped the traditional perception of ray tracing, transforming it into a viable alternative to GPU-based rasterization approaches as the default method for visualization. The presence of highly optimized and extensible ray tracing engines, like intel OSPRay~\cite{OSPRay} and the GPU accelerated raytracing, has significantly influenced recent research. Many visualization tools, like ParaView, VMD, VisIt, and MegaMol have adapted intel OSPRay as rendering backends. OSPRay~\cite{OSPRay} is a scalable, CPU-based ray tracing rendering library built on Embree\cite{Embree} and can visualize massive data as long as it fits the available CPU memory. Recently, MegaMol implemented RTXPkD~\cite{RTXpkd}; a GPU-based Particle Kd-Tree (Pkd) method using the OptiX GPU ray tracing framework. RTXPkD was able to render a dataset containing 1,56 billion spheres in 75 FPS on NVIDIA Quadro RTX 8000 with 48GB VRAM. Existing approaches can only visualize models that fit into the CPU and/or GPU memory, we overcome this limitation by interweaving the construction and the rendering. We propose a view-guided scene construction technique that constructs a fraction of the scene visible from a given viewpoint that can fit the dedicated video memory.} We adopt an RTX ray-tracing pipeline which results in rendering performance speed up by 2-3 orders of magnitude as compared to previous rasterization approaches.
%+++++++++++++++++++++++++++++++++++++++++++++++++++++++++++++++++++++++++++++++++++++++++++++++++++++
\vspace{-0.44cm}
%+++++++++++++++++++++++++++++++++++++++++++++++++++++++++++++++++++++++++++++++++++++++++++++++++++++++++++++
\paragraph*{\textbf{Parallel Rendering:}} 
Parallel rendering aims to improve the frame rate by dividing the workload between multiple renderers. Generally, two of the three rendering architectures classified by Molnar et al.~\cite{Molnar-1994-A-sorting-classification-of-parallel-rendering} have taken the most attention. The first category, so-called {\it sort-first}, involves dividing the screen into disjoint regions and each renderer is responsible for all computations that are related to its region. Each renderer accesses its own copy of the scene, making memory management inefficient. Another rendering architecture category, so-called {\it sort-last}, divides the scene's primitives rather than the screen. Renderers compute a full-screen image of their portion of the primitives and then submit these pixels to compositing processors that sort the rendered images in visibility order to produce the final image. This method can result in a very high data rate as the renderers operate independently until the compositing step. The {\it sort-last} strategy is suitable for applications that require interactive, high-quality rendering and efficient memory management. For more details, we refer the reader to~\cite{Molnar-1992-PixelFlow-High-Speed-Rendering-Using-Image-Composition,Molnar-1994-A-sorting-classification-of-parallel-rendering,Eilemann-2019-Parallel-Rendering-and-Large-Data-Visualization}. Our rendering algorithm uses the {\it sort-last} approach. 
Zellmann et al.~\cite{Zellmann-2020-Finding-Efficient-Spatial-Distributions-for-Massively-Instanced-3-d-Models} investigated how data could be efficiently partitioned across the parallel GPUs using a ray tracing-based renderer. They define the properties of good primitives partitioning; first, {even data distribution}, to make good use of the GPU memory and compute resources. Second, {even spatial distribution} because irregular spatial distribution may cause load imbalance and subsequently inadequate rendering performance. % as the ray-tracers bin rays to the spatial domain. 
 In addition, {minimize object overlap and replication} because in both these cases, the ray tracer would intersect the primitive more than once. Our method uses a spatial partitioning scheme to distribute space equally among cells. Each primitive in the scene belongs to only one cell, so no primitive is replicated.  Parallel rendering usually utilizes multiple GPUs to accelerate the rendering; however, our rendering algorithm is executed in a single GPU that runs multiple rendering threads in parallel through a compute shader.% and NVIDIA's \qcrFont{GLSL\_EXT\_ray\_query} extension. 

%=======================
\section{Technical Overview}
\label{sec:Technical-Overview}

\begin{figure*}[t]
    \centering
    \includegraphics[width=\linewidth]{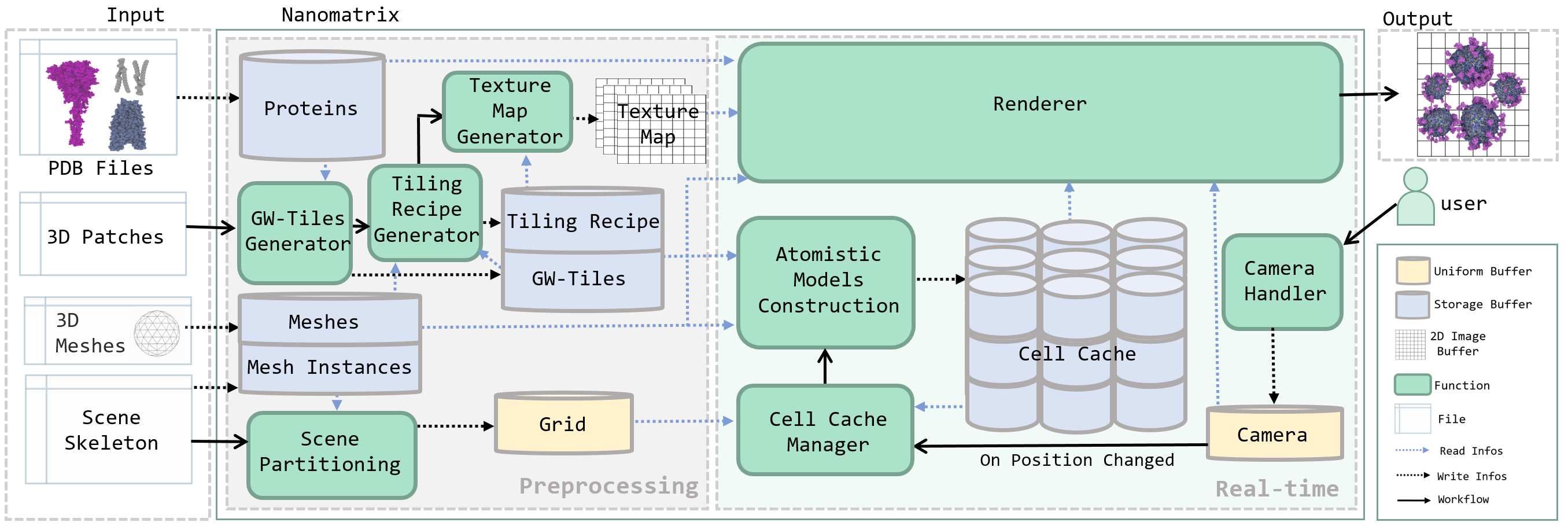}
    \caption{Nanomatrix -- the scalable construction algorithm. Based on the input structure and building rules, in the preprocessing step rectangular geometry-based or image-based aperiodic patches are generated. For representing a volume of molecules, box tiles are generated. In real-time rendering, these tiles populate the scene with the appropriate level of detail depending on the distance from the camera. The scene is then rendered using RTX ray tracing so that the population and rendering fully utilize separate computational units on the graphics hardware.}
   \label{fig:Scalable_Construction_Algo}
\end{figure*}
%=====================================================================

%\begin{figure}[t]
 %  \centering
%    \includegraphics[width=0.8\linewidth]{figures/moleculesVsMesoscaleModel.PNG}
 %   \caption{Protein molecule Vs Mesoscale Model\cite{goodsell-2015-rcsb}}
%   \label{fig:InstanceVsMesoscaleModel}
%\end{figure}

%\annot{motivation}
None of the currently available visualization methods is able to visualize large macro-molecular structures such as the red blood cell (RBC) in full atomistic detail or multiple instances of viral and bacterial ultrastructures. These structures are simply too large to fit into the CPU and/or GPU memory along with associated acceleration structures. At the same time, at any moment, only a small fraction of such a huge dataset can be seen. Therefore, a natural memory-saving and acceleration strategy is to store only those parts of the model in the memory that are visible from the current viewpoint. One solution is to use an out-of-core approach and stream the model from the disk to the core memory. However, instead, we are using an exclusive in-core approach and generating the model on the fly, where the camera triggers the process of generation. Our approach operates under the following condition: Biological structures that are far away are enclosed by their molecular envelope, such as a lipid bilayer for example. Far-away structures are never cut open or clipped in the middle, this happens only when the camera is close to a particular structure, only then the envelope can open to see the atomistic detail inside. 

%The goal of our approach is to visualize biological models that are too large to fit into GPU memory. A typical example would an atomistic model of a Red Blood Cell or many instances of viral and bacterial ultrastructure.
\begin{figure}[b]
    \centering
    \includegraphics[width=\linewidth]{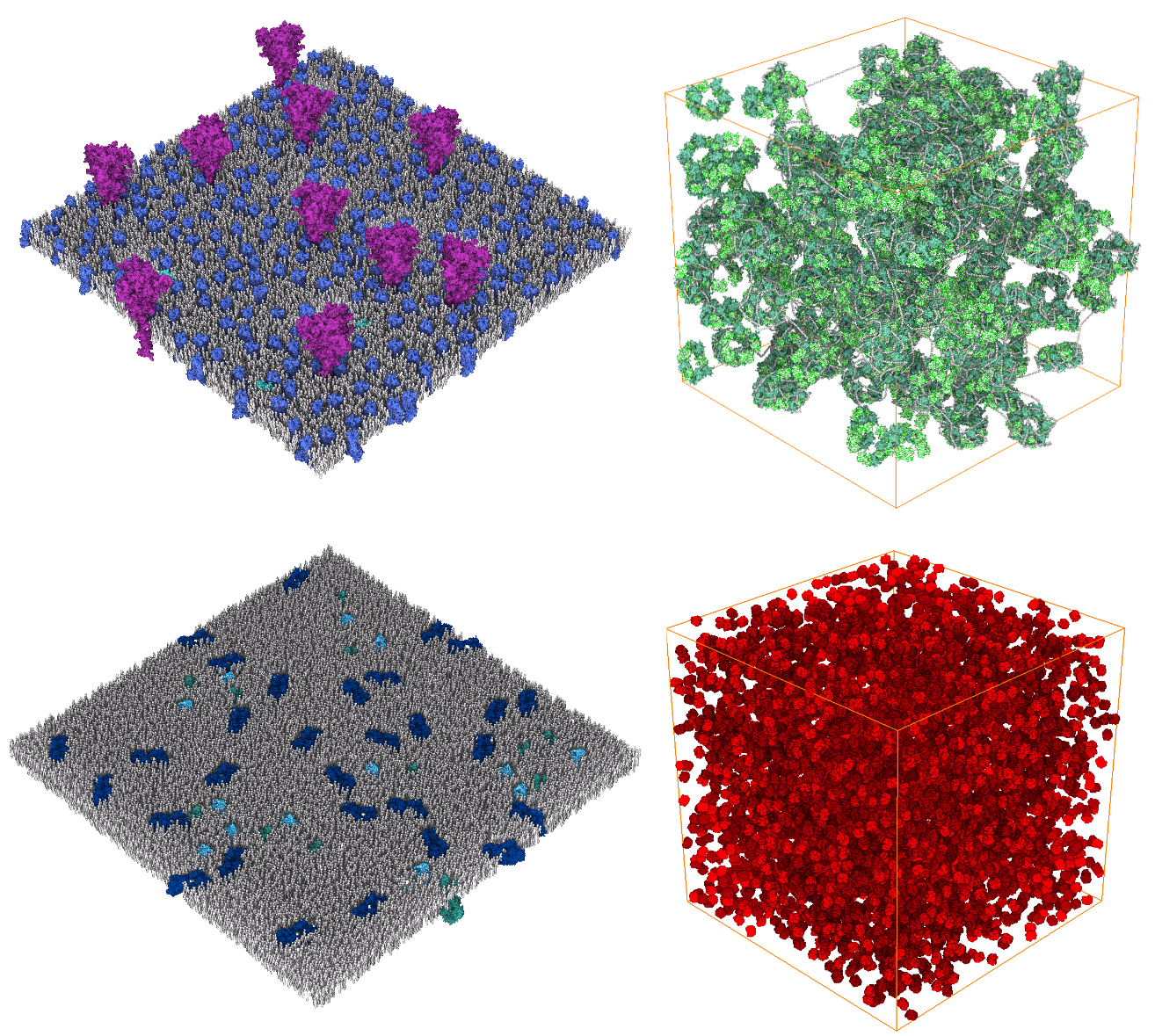}
    \caption{Illustration of a rectangle-based patch (l) and box-based patch (r) for SARS-CoV-2 (top row) and RBC (bottom row).}
   \label{fig:Patches}
\end{figure}
%\annot{Paper terminology}
Before we dive into technical detail, we clarify the terminology of the individual components involved in our approach (\autoref{fig:Scalable_Construction_Algo}). Our Nanomatrix approach generates geometry at atomistic detail of the {\it potentially visible} portions of the view. By {\it potentially visible} we mean structures that are either (1) close to the camera, so that the atomistic detail becomes discernible or (2) there is a clear view of these structures unobstructed by other densely packed molecules. Our approach also generates image texture representation for those structures that are potentially visible but are so far away from the camera that their atomistic detail is no longer perceivable. To realize this, four elements are expected as input. First, the structure of all types of molecules that will be populated in the scene is stored in \textbf{PDB files}. These files are available at the {\it Protein Data Bank}  (\href{https://www.wwpdb.org/}{wwpdb.org}). %(\href{https://urldefense.com/v3/__https://www.wwpdb.org/*7D*7Bwwpdb.org__;JSU!!Nmw4Hv0!x5BQGCPTJHPfAMIGRqtFmrG128zSrXSBB2WjAftN0PrNtrGRQJ9upkp5rFHAiA3uINkXAbDONC_MwGAUCL3uZCS9caO8lwNCjg$ }). 
The second input type is \rh{the \textbf{3D patch}, which }is a small, collision-free 3D biological model that is constructed based on \rh{biological} %\textbf{rules} using MesoCraft~\cite{Nguyen2021} tool.
{\it rules}~\cite{Mesocraft}, which define the concentration of various molecules in the patch and the principles that spatially characterize their spatial relations. \rh{These can be created in multiple ways, e.g. using the MesoCraft tool~\cite{Mesocraft} or some other alternative}. Our approach uses two types of patches: the \textbf{rectangle-based patch} and the \textbf{box-based patch}, both are illustrated in \autoref{fig:Patches}. The third input is the \textbf{3D mesh} which defines the geometry of a given biological compartment. The scene may contain several copies of \rh{various biological structures. The} information that is needed to instantiate the given meshes is given in the \textbf{scene skeleton}, \rh{which is defined as an external file. }

Once the input files are given, the pre-processing phase starts preparing the necessary components for the construction and rendering phases. As we aim to visualize models that do not fit into the memory, we need to partition the space and keep only potentially visible parts of the scene in the memory. The \textbf{scene partitioning} divides the 3D space into small non-overlapping \textbf{cells} of identical \rh{but customizable} sizes, along each axis that all together form the regular scene \textbf{grid}. Unless explicitly stated in the remainder of the paper, we refer to the {\it cells} as the elements of the scene grid as a technical term. Our visualization is concerned with biological cells, but if this term occurs in that context, it will be explicitly denoted as being a {\it biological \rh{entity}}. Also, the {\it cell} is an element of global scene partitioning, in contrast to the {\it box} or {\it rectangle tiles}, which are elements forming the detail of a particular biological entity. Each cell is associated with an {\it index} which defines the cell location $(i,j,k)$ within the grid. At any moment, only a small group of cells will be populated. We call this group of cells the \textbf{active cells}. To identify them, the camera's location within the grid is obtained first, which represents the \textbf{central cell}. This cell, together with its neighboring cells, represents the active cells. The number of active cells depends on the size of \textbf{activation window}, which indicates the number of central cell's neighbors that should be considered active. Each active cell points to a \textbf{cell cache} which is a GPU storage buffer that is readily available to be filled with geometric instances forming the structural information of a biological entity.

%Additional important components that are essential for the proposed construction approach are the \textbf{GW-tiles} and \textbf{seamless tiling recipe} generated in the prepossessing phase. The 3D patch is used to generate a set of Geometric Wang Tiles (\textbf{GW-tiles}) as described in Klein et al.~\cite{Tobias-2018-Instant-Construction}.
%\rh{\annot{Difference between Geometric Wang Tiles and the original Wang Tiles (R3.3)}
\rh{Once a cell is selected to be active, Nanomatrix generates the cell contents on the fly. The proposed construction approach uses the {\it Wang tiles} concept~ \cite{Wang1961} for populating the cells with geometries from a set of geometric tiles. While the original {\it Wang tiles} algorithm covers an infinite plane with a virtual texture generated from a small set of image tiles.}  We introduce a novel concept of aperiodic geometric tiles \rh{({\it geometric texture})}, denoted as \emph{Geometric Wang tiles} or \textbf{GW-tiles} for short. \rh{While the image tile consists of an array of pixels, geometric tiles consist of an array of instances each associated with a transformation matrix. Every Wang tile is defined as a square with color-encoded edges. These colors restrict how the tiles can be placed during the tiling process to form a seamless tiling.} In the pre-processing step, a description of such arrangement is created and denoted as \textbf{seamless tiling recipe \rh{(TR)}}. This recipe is described \rh{as} a 2D table of pointers that refer to one of the generated {GW-tiles}.

For \rh{biological} structures that are far away from the camera \rh{and} cannot be observed closely, the detail is represented by an image texture, instead of a geometry texture. Therefore, once a GW-tile is obtained, its geometry is used to generate the \rh{\textbf{packed texture map} where we pack the Wang tiles' textures into a single texture map. During the run time, for each texture request, we first determine which tile that fragment is located at, based on the position of its $uv$ texture coordinates and the tile recipe. We then compute the relative offset of $uv$ within that tile and fetch the corresponding texel from the packed texture map. The \rh{resulting texture rendering} is composed of diffuse, normal, \rh{depth}, and ambient occlusion textures which} can be used in the deferred shading step when mapped on a mesh. 

During the rendering, when the camera moves to a new location \rh{and if necessary} the \textbf{cell cache manager} will update the active cells, as well as the pointers to the cell cache, and then submit the cells that need to be generated to the  \textbf{atomistic models construction}. This procedure populates the proteins in the cells using {GW-tiles} and {seamless tiling recipe}. The constructed scene is rendered in two passes. In the first pass, each active cell computes a full-screen image of their portion of the atomistic models using hardware-accelerated ray tracing that is available on GPUs. The resulting images are then composited in the second rendering pass to form the finally rendered image. During the compositing rendering pass, if the ray hits an atom in one of the active cells' image buffers, the closest hit is used to assign the final color and the ambient occlusion value is computed. Otherwise, the {\it \rh{packed} texture map} is used to texture meshes in the scene.

%=======================
\section{Pre-processing Phase}
\label{sec:Pre-processing-Phase}
%\annot{Why do we need to partition the scene? Highlight the motivation here}
%To support the fast generation of the model, we need to prepare and preprocess some data. More specifically,
In the pre-processing step before the rendering is started, we partition the scene and fill the scene with mesh instances. We also prepare the geometry for all meshes and prepare and upload buffered data to the GPU. 

\subsection{\rh{Virtual} Scene Partitioning}
%\annot{3D meshes initiating}
%\annot{What is mesh; how is it loaded; several different types with different positions and rotations.}
The entire scene is partitioned into several cells which are filled during real-time rendering with structures on demand. These cells are organized in a grid that covers the entire scene. To create the grid, the first task is to define the axis-aligned bounding box (AABB) that tightly encloses the object distribution in the scene. These objects represent biological structures, such as biological cells, viral particles, bacteria, or organelles, which come in different sizes, and shapes. We use 3D meshes to define the boundaries that separate the internal parts of these structures from the outside environment. As the scene may consist of several copies of the same structure, the scene skeleton file contains information needed to instantiate the given meshes in the scene. This file contains a list of mesh instances and for each instance, the {\it mesh-id, position}, and  {\it rotation} are provided. The meshes and scene's skeleton are together used to estimate the scene's {grid} AABB.

%\annot{Virtual partitioning only, no grid stored in memory. The scene is subdivided into cells.}
%\annot{motivation for choosing uniform space (R3.2)
\rh{ Since we are targeting to render large scenes, partitioning the space into cells will yield a large number of cells and correspondingly large memory requirements. We avoid that by using the uniform space subdivision scheme. } \newText{ For more detail check the supplementary material.}

\subsection{Tiles Preparation}
\label{sec:Tiles-Preparation}
\begin{table}[t]
\centering
\caption{\newText{Symbols used in this paper}}
\scalebox{0.8}{
{\begin{tabular}{p{0.20\linewidth}p{0.70\linewidth}}
\hline
Symbol & Explanation\\
\hline
%\textbf{$|...|$}& number of elements in a set \\
\textbf{$H$}& a set of mesh instances in the scene  \newText{$H =\{h\}$} \\ %$h \in H$ \\
\textbf{$T$}& a set of triangles in the mesh  \newText{$T =\{t\}$} \\ 
\textbf{$C$}& a set of active cells  \newText{$C =\{c\}$} \\
\newText{\textbf{$c.T,$ $c.H$}}& \newText{ a set of triangles/mesh instances that intersect the active cell $c$ (where $c.T \subset T$ and $c.H \subset H$)}\\
\textbf{$R$}& a set of geometric rectangular tiles \newText{$R =\{r\}$} \\ 
\textbf{$B$}& a set of geometric box tiles  \newText{$B =\{b\}$} \\
\textbf{$\hat{r}, \hat{b}$} &  a representative tile where  $\hat{r} \in R$ and $ \hat{b} \in B$. \\
\newText{\textbf{$\hat{c}$}} &  \newText{a representative cell.} \\
\textbf{$m$}& a molecular instance\\% within a geometric Wang tile $l$ or a box tile $b$\\
\newText{\textbf{$m_{uv}, m_{xyz},$ $ m_{\mu\gamma}, m_{\mu\gamma\lambda}$}}& \newText{the position of molecular instance in texture/world or tile coordinates. We use subscript $uv$, $xyz$ and
${\mu\gamma\lambda}$ to differentiate between the texture, world and tile coordinates, respectively}.\\
%\newText{\textbf{$f_{uv}, f_{\mu\gamma}$}}& \newText{the position of a fragment in texture or tile coordinates}.\\
\textbf{$TR$}& tiling recipe; which represent a 2D array of pointers where each element of this array refer to a rectangular tile $r \in R$. Conceptually, it refers to a pre-designed plan that shows how the Wang tiles can be placed next to each other such that the colors of shared edges are matched. \\
\textbf{$TR[i,j]$} & a pointer to a tile that is located at the 2D index $[i,j]$ in TR\\
\textbf{$TR_{ij}$} & 2D indices $(i,j)$ within TR\\[0.1cm]
%\textbf{$rep$}& replication area; which is a 2D/3D array  that represent a subset from $TR$ that mapped to a particular triangle $t$.\\
\newText{\textbf{$rep^{R},$     $rep^{B}$}} & replication area; %tiling window ???% defines a window by which we use a subset of the TR for covering the entire triangle/cell
which is a 2D/3D array of tiles that are used for covering the entire triangle/cell. These tiles are selected based on $TR$ in the case of triangle tiling while they are randomly selected in the case of cell tiling. \newText{We use superscript $R$ and $B$ to differentiate between the replication area of rectangular tiles and box tiles, respectively.}\\[0.1cm]
\newText{\textbf{$rep^{R}_{ij},$ $rep^{B}_{ijk}$ }} & 2D/3D indices within the replication area.\\[0.1cm]
\newText{\textbf{${N}^{R},{N}^{B}$}}& the maximum number of molecular instances that can be found in a rectangular/box tile.\\
\newText{ \textbf{$.size_{uv},$ $.size_{xyz}$}} & \newText{2D/3D vector that represents the length, width and height of the associated tile, triangle or cell in texture/world coordinate.}\\
%\newText{ \textbf{$\text{size}(...)_{uv},$ $\text{size}(...)_{xyz}$}} & \newText{function that returns 2D/3D vector that represents the length, width and height of the given tile, triangle or cell in texture/world coordinate .}\\
%\newText{ \textbf{$\text{min}(...)_{uv},$ $\text{min}(...)_{xyz}$}} & \newText{function that returns the minimum coordinate of the axis-aligned bounding box of the given tile, triangle or cell in texture/world coordinate .}\\
\newText{ {$.min_{uv},$ $.min_{xyz}$}} & \newText{the minimum coordinate of the axis-aligned bounding box of the associated tile, triangle or cell in texture/world coordinate .}\\
\newText{{$.dim$}} & \newText{the dimensions of the associated multi-dimensional array.}\\
\textbf{$|...|$}& number of elements in the given set \newText{or multi-dimensional array}. \\
%($Lmax\times rep_u\times rep_v)$
% {N}^{L} \times |rep^{L}|
\hline
\end{tabular}}}
\end{table}

%\annot{Why tiles-based and rule-based approach?}
Our strategy for filling a certain 3D volume with particular molecular instances is by filling it with a limited collision-free set of 3D tiles that are prepared and stored during the pre-processing step, intended for use during real-time rendering. With such a construction approach, we overcome the computational load needed for populating huge numbers of elements on the fly. Moreover, this construction minimizes the collisions of populated elements.

\begin{figure}[t]
    \centering
    \includegraphics[width=0.8\linewidth]{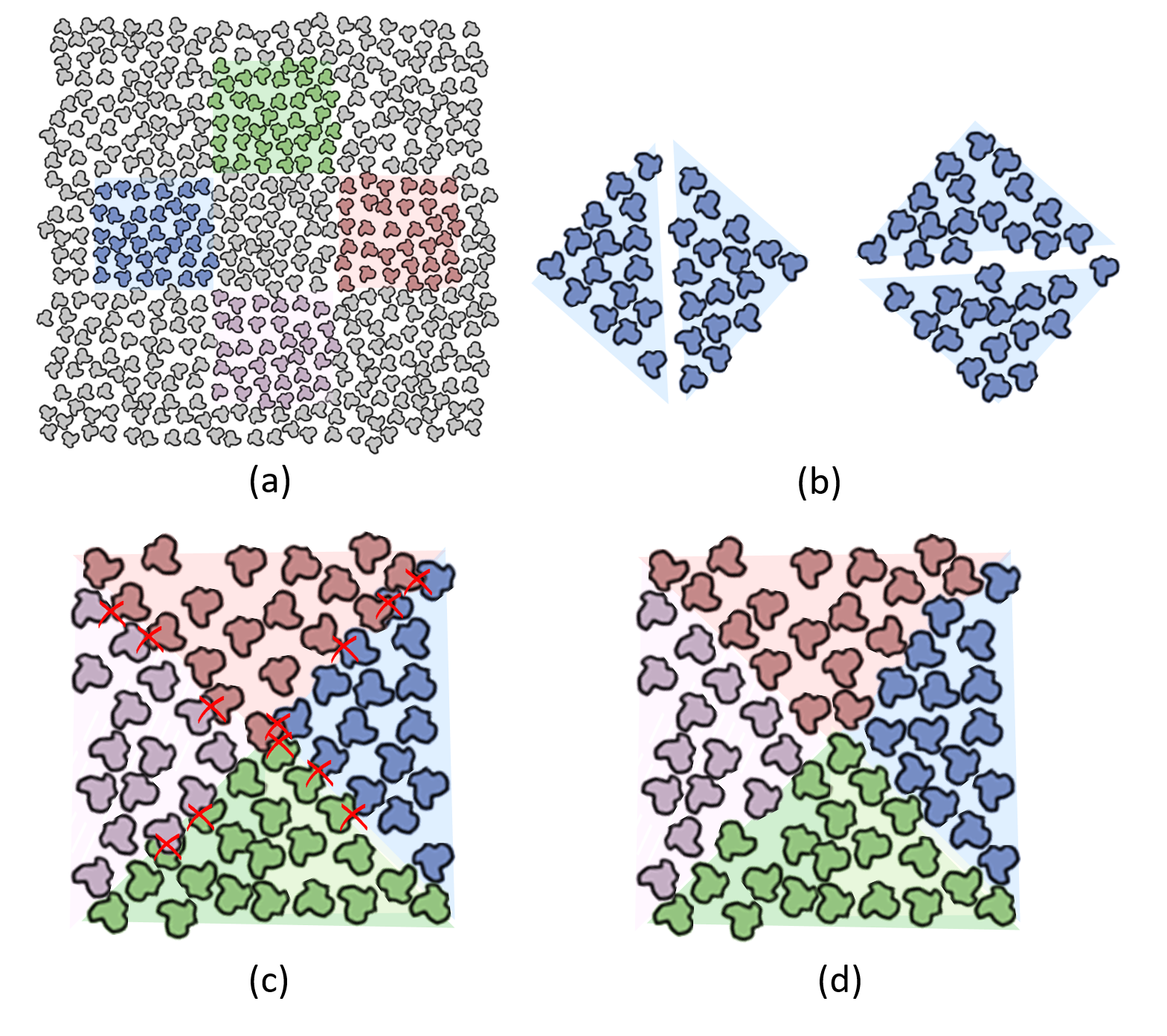}

    \caption{\newText{Illustration of the construction process of a single geometric Wang tile}. (a) four randomly chosen initial base patches inside rule-based geometry patch each of them is associated with a color. (b) the four colored base patches are further subdivided into four triangular sub-patches to be used for horizontal and vertical edges. (c) a Wang tile is created by combining four triangular sub-patches, collisions need to be resolved between the stitched triangular sub-patches.  (d) the Wang tile has been further processed to resolve the collision.}
   \label{fig:Wang-Tile-Synthesis}
\end{figure}

\subsubsection{\rh{3D Patches Generation}}
First, a 3D patch populated with molecules is generated. For this purpose, we use the MesoCraft tool that was designed for generating biological assemblies based on the simple geometrical rules that define relations between elements, and molecular structures in our case~\cite{Mesocraft}. The rules description is out of the scope of this paper, however, it is important to mention that MesoCraft integrates the collision handling algorithm. Therefore, the resulting patch is collision-free. We work with two different patches; rectangle-based and box-based. A rectangle-based patch is formed by molecular geometry populated on a rectangle. The molecular distribution is with a reference to a plane and is later used for populating a membrane. The second type, a box-based patch, is formed by populating molecular geometry inside a box (distribution in a unit of volume) and is typically used for populating soluble proteins inside structures. We create box patches with the dimension of $100 \times 100 \times 100$~nm filled by approximately 1,000 protein structures. The comparison of both patch types can be seen in~\autoref{fig:Patches}. 
%After replicating several \rh{molecular} instances next to each other, our experiments show that the seams, resulting from periodic tiling, are not noticeable for box tiling. The reason is that this tiling is only used for filling an internal part of a biological structure. After the camera penetrates inside the structure, due to the densely populated environment, the user is immersed among multiple structures, which leaves a limited possibility to identify the seams. Therefore, we do not additionally process these box-shaped patches. Rectangle-based patches form membranes that the user can see from the outside. For this reason, the rectangle-based patches need to be processed further, to guarantee a seamless geometry, see below.

\begin{figure*}[t]
    \centering
    \includegraphics[width=1.0\linewidth]{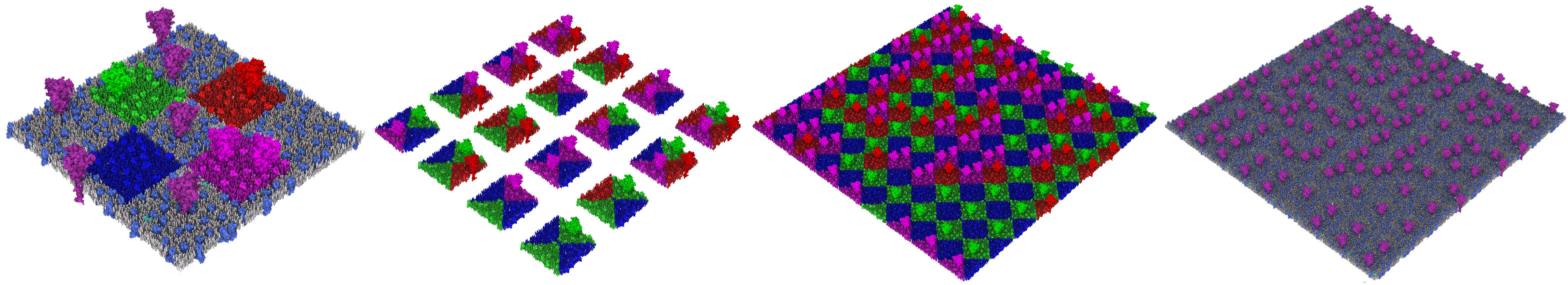}
    \caption{Illustration of the tiling algorithm \newText{that shows the application and visualization of the geometric Wang tiles of SARS-CoV-2 model}. From left: four randomly chosen initial \rh{base} patches inside rule-based geometry patch, 16 \rh{Geometric} Wang Tiles, the population of the tiles using a tile recipe rendered with and without Wang Tiles encoding.}
   \label{fig:Tiling}
\end{figure*}
\begin{figure}[t]
    \centering
    \includegraphics[width=1.0\linewidth]{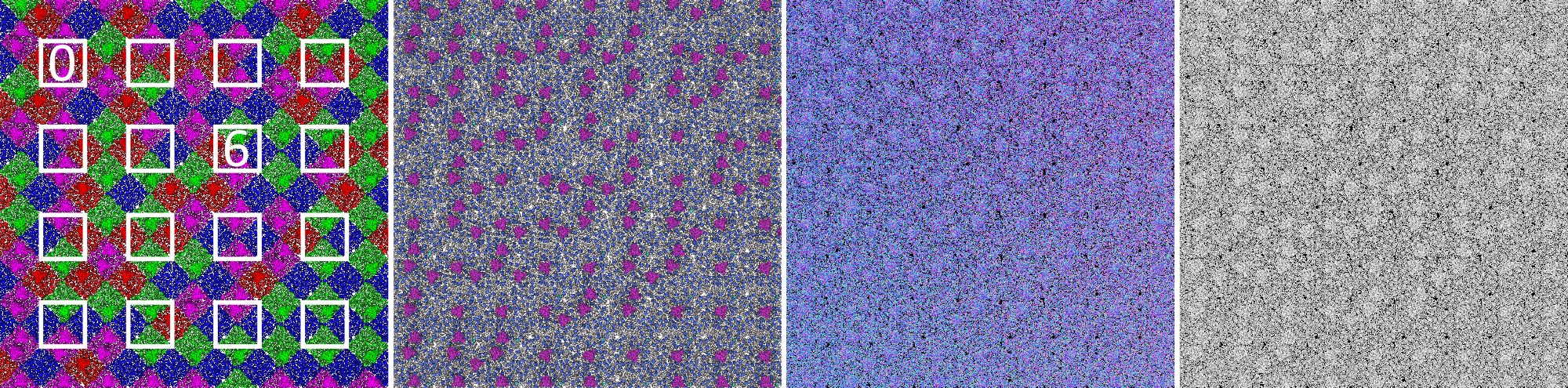}
    \caption{Illustration of the textures obtained from the geometry tiles. From the left: texture generated by Wang tiling with highlighted 16 base tiles; Diffuse texture; Normal texture, Ambient-Occlusion texture. }
   \label{fig:Textures}
\end{figure}
\subsubsection{\rh{Generating Rectangular Wang Tiles}}
\label{section:wang-tiles}
\rh{Placing these patches directly next to each other to populate the mesh will potentially create an overlap on the edges of adjacent patches. LipidWrapper~\cite{LipidWrapper} deletes overlapping molecular instances and fills the resulting holes with new instances. This is an expensive process which makes it unsuitable for the interactive scene population. Klein et al.~\cite{Tobias-2018-Instant-Construction} avoid that by applying the Wang tiling concept. The process of generating geometric Wang tiles is illustrated in~\rh{~\autoref{fig:Wang-Tile-Synthesis}}.} 
%\annot{Generating geometry tiles and tiles recipe}
%To generate seamless {\it geometry textures}, we implemented the Wang tiles concept that works with 16 base tiles. \rh{To generate These tiles, we follow the approach of Klein et al.~\cite{Tobias-2018-Instant-Construction}.} 
%The process is illustrated in\autoref{fig:Tiling}.
From the geometry rectangle-based patch that is collision-free (see \autoref{fig:Patches} (left)), four initial non-overlapping base patches \rh{$K$} are randomly selected and each of them is associated with a color \rh{(~\autoref{fig:Wang-Tile-Synthesis} (a))}. From the base patches, a set of tiles \newText{$R$} using the Wang tiles approach is created in the following way. Every base patch from \rh{$K$} is subdivided into four triangular sub-patches \rh{(~\autoref{fig:Wang-Tile-Synthesis} (b)) to be used for
horizontal and vertical edges.} A \rh{Geometric Wang} tile \newText{$r \in R$} is created by combining four triangular sub-patches from any of \rh{$K$} to form a rectangle \rh{(~\autoref{fig:Wang-Tile-Synthesis} (c))}. %Therefore, the collisions need to be solved if two neighboring triangular sub-patches come from different base patches.
\rh{The generated Wang tiles have to be further processed to resolve the collisions generated between the stitched triangular sub-patches (~\autoref{fig:Wang-Tile-Synthesis} (d)). }
 Later in the population phase, if two \rh{Wang} tiles are laid next to each other so that the triangular sub-patches on both sides of the shared edge are from the same base patch (are associated with the same color), they do not create a seam and there is no need to solve the collision. We work with basic 16 configurations of \rh{colors for Wang} tiles (can be seen in detail in \autoref{fig:Tiling}). However, more configurations can be taken into account. \rh{In general, if there are $|K|$ colors then there are $2\times |K|^2$ combinations of colors tiles~\cite{WangTilesforImageandTextureGeneration}. However, less number of tiles could still be used to create aperiodic tileset~\cite{An-aperiodic-set-of-11-Wang-tiles}. } The more configurations are used, the bigger the diversity of the generated result.

\paragraph*{\textbf{\rh{Rectangular Wang Tile Mapping}}}
%After the tile-set $L$ is created, for every molecular instance $m_i \in l$, where $l \in L$, its $uv$-texture coordinate $uv(m_i)$ within the rectangular tile $l$ is computed. This coordinate is within $[0,1]$ space and we call it the {\it tile texture coordinates}. 

After the \rh{geometric rectangular Wang} tile-set \newText{$R$} is created, we need to define a way for mapping a molecular instance $m \in \newText{r}$, where \newText{$r \in R$}, from the tile into the mesh. To achieve that, we need first to identify the location of that instance within the tile $m_{\mu\gamma}$. The $\mu\gamma$-coordinate is within $[0,1]$ space and we call it the {\it tile coordinates}. It can be computed by applying min-max normalization that rescales the instance position from the world coordinate ($m_{xyz})$ into $[0,1]$.% as it is shown in~\autoref{alg:assignTileCoordinats}. In the algorithm, we assume that the tile's instances are perpendicular to the \newText{$xy$}-plane. }

\rh{Klein et al.~\cite{Tobias-2018-Instant-Construction} \newText{propose per face tile mapping approach. They} use the quadrilateral coordinates to map the instance $m$ from Wang tile to the mesh quad. Therefore, their method required the mesh to be defined as an equally-sized quad-based surface which is a significant constraint as it can be difficult to create such a mesh for an arbitrary shape. In this work, we propose to rather use triangular mesh texture coordinates for the mapping, which enables us to map the Wang tiles onto triangle meshes of varying triangle sizes directly.}

Our method expects a triangular mesh to be given as input, where every triangle is associated with the texture coordinates as well. We expect that the mesh already contains texture parameterization and both {\it mesh $uv$-texture coordinates} are within $[0,1]$ range. The algorithm uses the texture coordinates for transforming molecular elements $m$ from tile coordinates to mesh texture coordinates and vice versa. If the mesh does not have a texture parameterization, a simple cube-map or spherical texture parameterization can be applied and used. However, depending on the shape of the mesh, it might be non-trivial to create fully seamless texturing. A seam in texture parameterization would result in a visible seam. The scene can contain multiple instances of different meshes. For simplicity, in the rest of the paper, we refer to a single mesh. However, the method works in the same way for every mesh that has a texture parameterization.

As we are aiming to run the population phase in parallel, we use the largest triangle $t_{big}$ of the mesh $T$ to define the mapping from the mesh texture coordinate system into the tile coordinate system. This mapping is essential for thread-count estimation, \newText{as explained later in~\autoref{sec:Construction}. }%\autoref{sec:ThreadsEstimation}.}
 All the tiles in \newText{$R$} are of the same size. We compute the ratio between the size of $t_{big}$ and one representative tile \newText{$\hat{r}$}. The ratio represents the number of tiles needed for tiling to cover the entire area of triangle $t_{big}$ with a sequence of tiles in its plane \newText{which denotes the dimensions of \emph{Replication Area (rep)}. It is} calculated as,
 \begin{equation}
\newText{rep^{R}.dim = \lceil\frac{t_{big}.size_{xy}}{\hat{r}.size_{xy}}  \rceil}
\end{equation}
 Moreover, as the mesh is texture parameterized, the mapping \newText{($\hat{r}.size_{uv}$)} that represents the size of a single tile in the texture coordinate space associated with the mesh, is computed by dividing the size of the triangle $t_{big}$ in texture coordinate over the ratio \newText{($rep^{R}.dim$)}. %(see~\autoref{alg:defineMappingAndTilesRecipeSize}).}

\rh{For any triangle in the mesh, we can now cover the triangle with Wang tiles such that the tiles with the same edge color are placed side by side to create seamless tiling within that triangle. To populate all triangles in parallel, we need to know in advance which part of the mesh is covered by which Wang tile. Therefore,} we generate a \rh{\it tiling recipe} $(TR)$ from the tile set \newText{$R$} that is associated with the mesh. A \rh{\it tiling recipe} is a 2D array that contains indices of Wang tiles from \newText{$R$}  (in our case a value from the range $[0..15]$ as we work with 16 tiles) and represents a lookup table when populating the respective part of the mesh during real-time rendering. \rh{The tiling recipe is designed to be large enough to cover the full texture coordinate space $[0,1]$. }

\rh{Its size can be computed as the inverse of \newText{$\hat{r}.size_{uv}$}. Once the size of tiles recipe $TR$, which is a 2D array, for the entire mesh is defined, it} is filled with the indices of tiles from \newText{$R$} by the Wang tiling generator. This structure is prepared for later sampling to determine a tile at an arbitrary texture coordinate that belongs to the mesh. By dividing a $uv$ mesh texture coordinate by \newText{$\hat{r}.size_{uv}$}, we get a two-dimensional index into $TR$. The other way around, if we multiply $TR$ \rh{index} by \newText{$\hat{r}.size_{uv}$}, we get \rh{the position} in the mesh texture. %To this point, previously described computations are performed only once. The description of the iterative algorithm that populates an active cell $C$ \rh{described in~\autoref{sec:Construction}}.
\subsubsection{\rh{Generating Box Tiles}}
Biological models are typically packed with proteins, nucleic acids, and other molecules. Soluble components fill the inner part of the biological compartments. The population task is to distribute the soluble ingredients spatially while avoiding overlaps. CellPack~\cite{cellPACK} positions the soluble ingredients sequentially one by one while avoiding collisions which is an expensive process that makes it unsuitable for the interactive scene population. Instant construction~\cite{Tobias-2018-Instant-Construction} uses parallel processing to increase the performance through consecutive steps where the space is first filled with instances through parallel threads while ignoring the overlaps. Then the overlaps are detected in parallel and resolved. Due to the density of biological models, the collision-resolving process may not converge and has to be terminated after a certain number of iterations. Instead, we propose to populate the soluble components by filling the space in parallel with collision-free 3D tiles which eliminate the need for real-time collision handling.
\rh{Wang cube~\cite{Culik-1995-An-Aperiodic-Set-of-Wang-Cubes} is an obvious generalization of Wang tiles to three dimensions. A Wang cube is defined as a cube with colored faces. Although it is possible to generate Wang-cube-tiles from the box-based
patch to later place them in the space based on Wang's tiling concept to create a seamless 3D geometric texture. We use the box patch as a box tile to populate the scene space directly. We do not use a set of Wang cubes because we do not see a necessity to do that in our target scenes.}
After replicating several \rh{molecular} instances next to each other, our experiments show that the seams, resulting from periodic tiling, are not noticeable for box tiling. The reason is that this tiling is only used for filling the internal part of a biological structure. After the camera penetrates inside the structure, due to the densely populated environment, the user is immersed among multiple structures, which leaves a limited possibility to identify the seams. \rh{Unlike rectangle-based patches, which form membranes that the user can see from the outside, box-shaped patches are not visible from outside, only when penetrating the biological structure with the camera. Therefore, we do not additionally process the box-shaped patches and we use them directly to tile the 3D space}. 
%For this reason, the rectangle-based patches need to be processed further, to guarantee a seamless geometry.
%\rh{However, our approach can be easily extended to employ the Wang cube concept which will guarantee that the 3D structures can be seamlessly stitched.}
\paragraph*{\textbf{\rh{Box Tile Mapping}}}
We need to define a way for mapping a molecular instance $m \in b$, where $b \in B$, from a box-tile into the space. For that, we need to identify the location of that instance within the box-tile $m_{\mu\gamma\lambda}$.  Again, the tile coordinate can be computed by applying min-max normalization that rescales the instance position from the world coordinate
$(m_{xyz})$ into $[0, 1]$. %  as it is shown in~\autoref{alg:assignTileCoordinats}. %As all the box tiles in $B$ are of the same size, the ratio between the size of the grid cells $cell.size_{xyz}$ and one representative tile \rh{$\hat{b}$} is computed. This ratio represents the number of box tiles $(rep_x, rep_y, rep_z)$ needed for tiling to cover the entire area of the cell which is calculated as,}
To fill a cell with box tiles, we need to compute the ratio between the size of the grid cells \newText{$\hat{c}.size_{xyz}$}
and the box tile size. As all the box tiles in $B$ are of the same size, only one representative tile \rh{$\hat{b}$} is used to compute this ratio,
\begin{equation}
\newText{rep^{B}.dim = \lceil\frac{\hat{c}.size_{xyz}}{\hat{b}.size_{xyz}}  \rceil }
\end{equation}
\rh{It represents the number of box tiles needed for tiling to cover the entire area of the cell.}

\subsubsection{\rh{Packed Texture Map Generation}}
%\annot{Generating 2D texture tiles}
Scene structures can be viewed from a large distance, where the atomistic detail would gradually result in seeing variations of colors on a mesh surface, instead of recognizing any detailed geometry. Therefore, when the biological entity is far away, it is covered with an image texture instead of a GW-tile. However, we need to maintain correspondence between GW-tiles and the texture map. When zooming in, the rendering algorithm combines the texture mapping with atomistic detail and blends between them. To reduce the additional handling overhead during rendering, we create a single texture that contains all the tiles. Importantly, we need to correctly sample the values at the border of the tiles, which implies a special construction of the texture map.%, as described below.

We create a single texture map directly out of the GW-tiles. The texture map is not used in its continuum for texturing, it rather contains 16 texture tiles \rh{packed} for sampling, with a space between these tiles to secure correct sampling on the border of each tile.
We render a geometry patch from the top view along the $y$-axis such that the patch is aligned with the $xz$-plane. This patch consists overall of $9 \times 9$ GW-tiles which are placed in a seamless way between them. However, for sampling the texture map later in the real-time rendering stage, we only need $16$ tiles depicted in \autoref{fig:Textures}. The texture map is generated as follows: All $16$ GW-tiles are positioned by facing the $y$-axis into four rows and four columns with a gap between the tiles of the size of one tile in both $x$ and $z$ directions. Then, those gaps are filled in the first iteration in $x$ direction and in the second iteration in $z$ direction to complete a seamless Wang tile pattern. In the last iteration, a set of tiles is placed to create a border enclosing the previously placed tiles. Such a structure allows us to easily identify the coordinates of individual $16$ base tiles, which we use for sampling, based on their indices. For example, if the resulting texture map is rendered with the resolution of $1800 \times 1800$ pixels (meaning one tile is $200 \times 200$ pixels as this is $9 \times 9$ grid of tiles), the square in the texture (see \autoref{fig:Textures}) representing the tile with index $6$ is at the position $[600, 1000, 200, 200]px$ (top, left, width, height).

%\subsection{\newText{\sout{Estimating the Population Parallel Threads}}}
%\newText{\sout{For performance reasons in the population phase, there are many threads that populate the structure in parallel. A single population thread is responsible for processing a single molecular instance per triangle, in the case of rectangular tiling, and per cell, in the case of box tiling. We need to find a conservative estimation of threads to initialize before executing the population process. For that we identify a tile containing the maximal number of molecular instances and store this maximal number in {$N^{R}$ or $N^{B}$}, depending on whether it is a rectangular or a box tile. Then, in the rectangular tiling case, we identify how many tiles {$(rep_u, rep_v )$} are needed to cover the largest triangle in the scene {in its plane}. The total number of threads allocated for tiling each triangle is then the product {$Lmax\times rep_u \times rep_v$}. For the box tiling, we allocate the number of threads analogously, i.e., the number of instances in the box tile is multiplied by the number of tiles necessary for filling one entire cell ({$Bmax \times rep_x \times rep_y \times rep_z$}).}}
%=======================
\section{Cell Cache Management}
\label{sec:Cell-Cache-Management}

\begin{figure}[t]
    \centering
    \includegraphics[width=\linewidth]{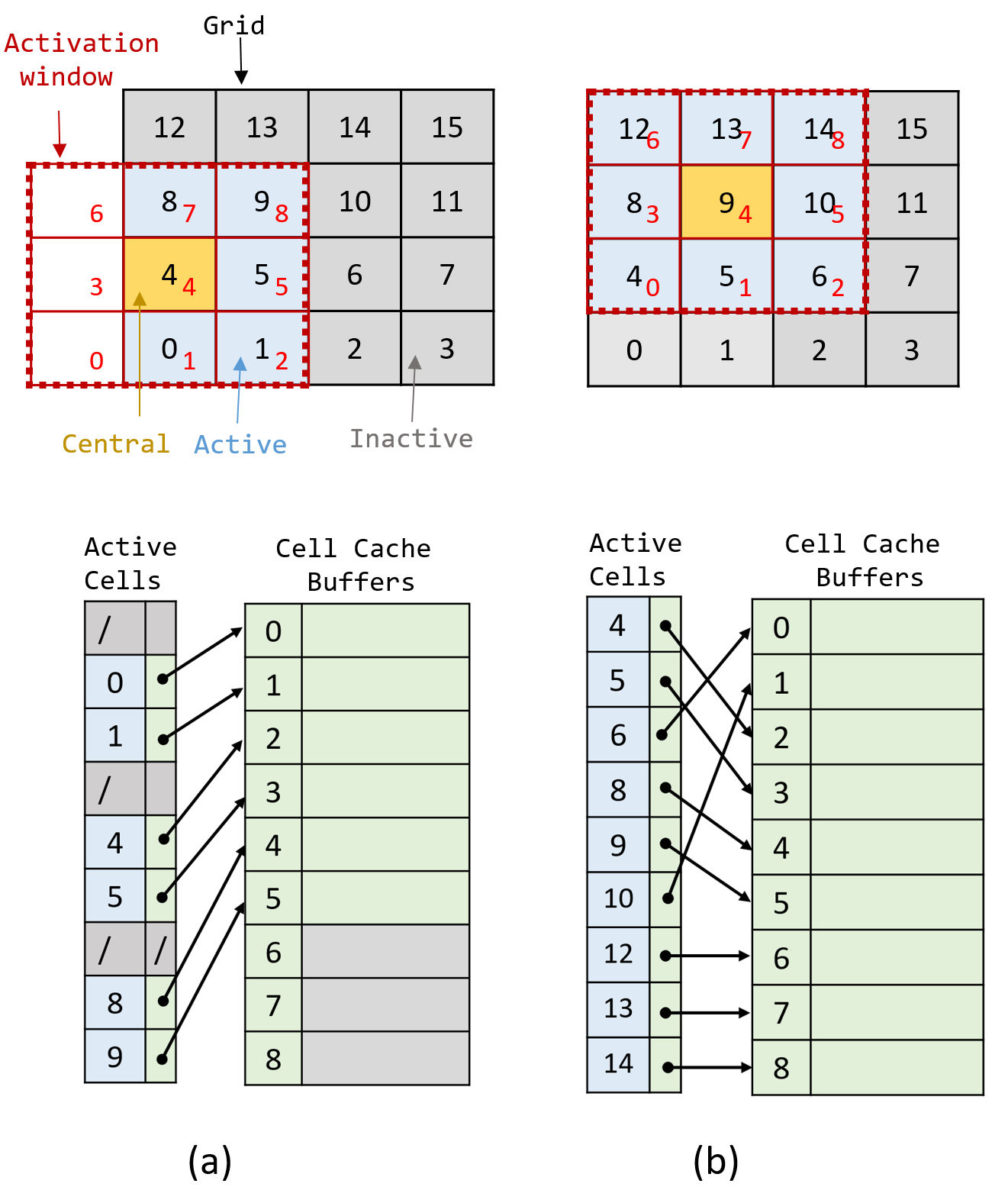}
    \caption{Illustrating the indexing-based algorithm for cache buffers dynamic allocation. In this 2D example, the scene grid contains $16$ cells and the number of active cells is $9$. (a) and (b) show two different central cells, where in each case the central cell is highlighted with the yellow color, while the active cells are with blue color.}
   \label{fig:indexing-based-algorithm}
\end{figure}

%\subsection{Active cells}
Only a portion of the scene geometry content is available in the memory at any given time during the real-time rendering stage. We achieve this using uniform space partitioning into cells. Next, we need to identify which cells among the scene's cells should be visualized and stored in the memory. We denote this scene's cells as the {\it active cells}.

%\annot{Camera position}
In our viewpoint-guided approach, the camera position is used to identify which cells should be active. Therefore, we first define the {\it central active cell}, which is the cell enclosing the camera. \newText{Its (i,j,k) index within the scene grid $G$} can be \newText{calculated as,}
\begin{equation}
\newText{\textbf{c}_{ijk} = \lfloor \frac{V_{xyz}-G.min_{xyz}}{\hat{c}.size_{xyz}}\rfloor}
\end{equation}
where \newText{$V_{xyz}$} is the camera viewpoint. After the central cell is identified, the neighboring cells are obtained. Thanks to the regularity of the grid, the adjacent cells of the central cell are easy to locate. \rh{Selecting which neighboring cells should be activated could be done based only on the camera position and activate the closest neighbor cells to the central cell. Another approach is to select them based on both the camera position and direction by activating the closest neighbor cells to the central cell that intersects the view frustum. For simplicity, let us assume that we choose the first approach and we} activate only the \rh{first} closest neighbor to the central cell in each axis $i,j$, and $k$, so the size of our {\it activation window} is ($3 \times 3 \times 3$) which gives us $27$ active cells $C$. However, based on the computational resources and settings of the size of the cells, \rh{the size of the activation window} can be set to a larger number. Increasing the size will increase the rendering overhead because each cell is drawn in a separate draw call, thus a larger number of images will need to be rendered for the final scene compositing.

%\annot{cell caches} 
The size of the activation window specifies the number of cells that will be populated and rendered. Subsequently, it specifies the number of {\it cell cache buffers} that need to be prepared. The {\it cell cache} is a GPU storage buffer that is readily available for the active cell to be filled with populated instances. This memory buffer is pre-allocated to fit a relatively large number of instances. In the prepossessing step, we allocate 27 storage buffers that represent the cell cache. As our scene is continuously regenerated, we choose to allocate cell cache in advance and just fill and clear them in real-time to avoid the overhead that comes from the frequent memory allocation and deallocation.

%\annot{explanation}
%While navigating through the scene, some cells will leave the activation window and should be destroyed and some enter it and should be constructed. 
The cache manager controls the process of reusing deactivated cells by updating the pointers between the cell cache buffers and active cells and triggers an event that leads to the regeneration of the scene inside newly active cells. \autoref{fig:indexing-based-algorithm} shows an example that illustrates this algorithm on a 2D grid. In this 2D example, the scene's grid contains $16$ cells and the size of the activation window is $3\times3$. %This algorithm is executed on the CPU. 
In \autoref{fig:indexing-based-algorithm} (a), the camera is located in cell 4,  which becomes the central cell and the cells $0$, $1$, $4$, $5$, $8$, and $9$ are the neighbors and all of them are inside the activation window. These cells are the active ones and should be populated. Every active cell should occupy a cell cache to fill it later with molecular instances in the construction stage. Once a cell becomes active, an unoccupied cell cache will be reserved for this cell and a pointer will be created to link them. This cell will be added to the list of cells that will be submitted to the construction stage. If the camera moves to the cell $9$ as shown in \autoref{fig:indexing-based-algorithm} (b),  the cells $6$, $10$, $12$, $13$, and $14$ entered the activation window and need to be constructed while the cells $0$ and $1$ left the activation window, therefore, their pointers to the cell cache have been deleted. This makes these cache buffers available for other cells.  Cells $4$, $5$, $8$, and $9$ were populated previously, and they are still pointing to the same previous cell cache buffers. These pointer operations are important to avoid copying between buffers. In other words, if a cache has been assigned to a cell, it will be reserved for that cell as long as the cell is inside the activation window.

\begin{figure}[t]
    \centering
    \includegraphics[width=0.60\linewidth]{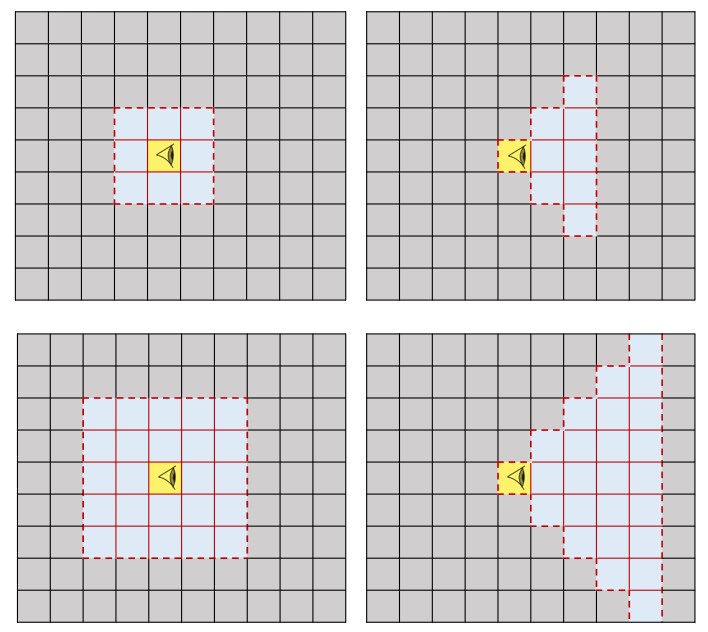}
    \caption{\rh{Illustration of different activation window layouts and sizes. In this 2D example, the central cell is highlighted with yellow while active cells with blue. The size of the activation widow is 9 cells on the top images and 25 on the bottom. In the left two images, the closest neighbor cells to the central cell are activated while on the right the closest neighbor cells to the central cell that intersect the frustum are activated.}}
   \label{fig:IllustrationaAtivationWindowLayouts}
\end{figure}

\rh{This cache management approach is applicable to any activation window layout and size. In our implementation, we have tested different possible layouts/sizes illustrated in~\autoref{fig:IllustrationaAtivationWindowLayouts}.}

\subsection{\rh{Occlusion Management}}
\rh{Biological models are densely packed with molecules. When the viewer enters such a dense world, an occlusion management technique has to be employed to visualize the model properly. In Nanomatrix, we employ a simple object-space clipping localized around the camera that discards elements according to their distance to the clipping geometric shape. This geometric shape, which is a sphere in our case,  specifies a region in object space that is influenced by the clipping. Besides managing active/inactive cells, the cache management invokes the instances visibility update when the camera moves to determine which instances within active cells should be shown or hidden. It tests the intersection between the instances bounding sphere and the clipping region to find out which ones lie inside the clipping region and sets them to be hidden. To accelerate the computation, the visibility test is done in a compute shader and for only the content of active cells that intersect our clipping geometry.}
%=======================
\section{Atomistic Models Construction}
\label{sec:Construction}
%\annot{Key properties of the algorithm:}
%\begin{itemize}
%  \item Combined approach based on texture coordinates and tiling.
%  \item No requirement on the mesh. Any triangular mesh with reasonable tessellation (without thin long triangles) will work.
%  \item No instance is created. The algorithm updates the pre-allocated buffer with instance type, position, and rotation.
%  \item GPU-based algorithm.
%  \item As good as the texture coordinates of the mesh are. The worse the texture coordinates, the worse the result is.
%\end{itemize}

In the pre-preprocessing phase, both rectangular-based and box-based geometry tiles are prepared. For their population, we implemented two distinct population methods. The first method is designed for the membrane population using rectangular \rh{Wang} tiles, the second \rh{method} is for the population of the inner matrix of the biological compartment using the box tiles. Both methods are described in this section.

%\annot{Why is membrane tackled differently than the rest?}
A membrane of a \rh{biological entity} is typically a thin envelope that consists of the lipid bilayer and membrane-bound proteins. %Because the membrane is targeted to be visible from a distance, it has to appear seamless so the observer cannot notice any artifacts caused by repetitive patterns. For the box-based geometry tile, while the viewers observe the inner part of the models, they are immersed in a heavily populated environment, where spotting any repetitive patterns is highly unlikely. 
For representing an overall shape of the \rh{biological entity}, we use a geometry mesh that can be created by 3D modeling tools or directly derived from biological measurements.
In the following, an algorithm that populates molecular instances along the mesh is described. \rh{The population is done for every activation cell independently. Therefore, in this section, we assume that the active cell $c \in C$ has been selected to be active and the task here is to populate this cell with geometries using rectangular tiles and box tiles}.

\newText{During the generation of the cell, many molecular instances have to be processed. We designed the construction algorithm to process and populate these instances independently which makes it suitable for parallel execution on GPU.} %Using the GPU for construction also has the advantage that the generated model already resides on the rendering device which eliminates the need for streaming the model from external storage.}
\newText{We estimate the number of parallel construction threads as the maximum number of molecules needed to cover the largest triangle in the mesh or a cell in the grid, depending on whether it is a membrane or solubles population.}

\subsection{Prepare Cell for Population}
%\annot{Intersected triangles are estimated}
\label{sec:TriangleIntersetion}
\rh{To ensure scalability, Nanomatrix is designed to prefer to recompute data over storing them. Therefore, no information about any cell in the scene grid is stored until it becomes active. Once a cell $c$ is selected to be active, the assigned cache for that cell $c.cache$ is cleared and some necessary information needs to be computed to prepare the cell to be filled with molecular instances.}

We need to populate the rectangular\rh{/box} tiles \rh{on/inside} the mesh. As the mesh can be of arbitrary size and shape, the active cell $c$ can be in three configurations: outside of a mesh, inside of a mesh, or intersecting a mesh. \rh{To populate it with molecular instances, we need to know first which mesh instances and triangles are intersecting that cell. We run an AABB-triangle collision algorithm in parallel to test for intersections between the active cell $c$ and all scene triangles and then store the intersected triangles $c.T$ and mesh instances $c.H$ in the cell's cache $c.cache$. If no intersection is found, that means the cell $c$ could be entirely inside or entirely outside the mesh, therefore, we need an in-out test. To achieve this, we find the closest triangle to cell $c.closestTriangle$ as the closest distance between the center of a mesh triangle and the center point of the active cell $c$. We find the closest triangle by using atomic operation in the same parallel compute shader.}
\begin{figure}[t]
    \centering
    \includegraphics[width=1.0\linewidth]{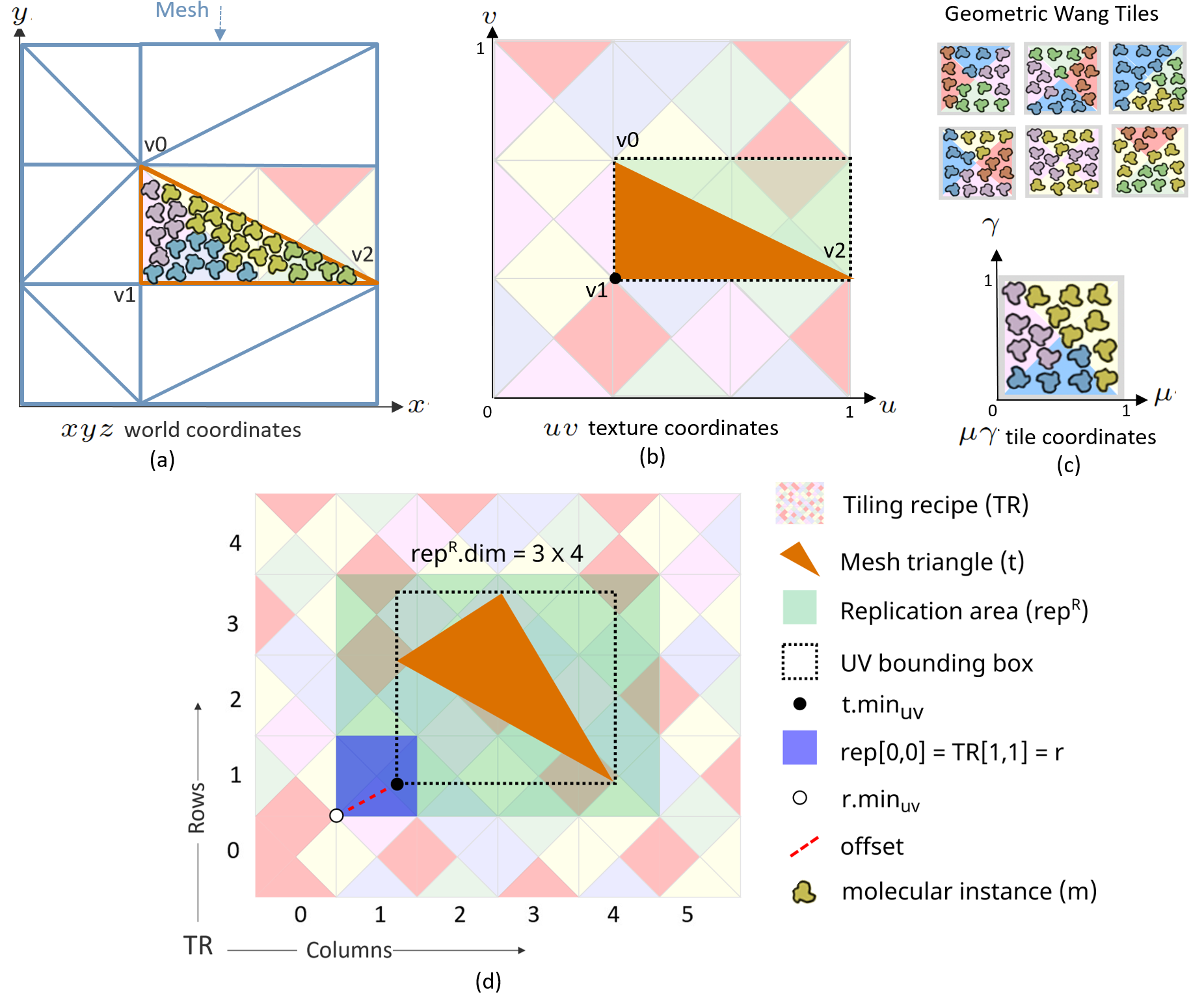}
    \caption{\newText{Illustration describes a simple example for membrane population at the top and a more general example at the bottom.} (a) we have a triangular mesh and we want to populate the triangle $t$ (highlighted with orange color) knowing its \emph{xyz} world coordinates and \emph{uv} texture coordinates. (b) \newText{projecting the triangle $t$ to the tiling recipe}. Based on the triangle's \emph{uv} texture coordinates, the replication area (\newText{$rep^{R}$}) that encloses the triangle $t$ is identified and then re-projected with its respective tiles onto the triangle in the 3D space. (c) the $\mu\gamma$ tile coordinates of a representative \newText{Wang} tile are illustrated. To populate a molecular instance $m$, we need to know its location in the world space $m_{xyz}$. To achieve that, we first map the instance from its tile coordinate $m_{\mu\gamma}$ to \emph{uv} texture coordinate $m_{uv}$ and then we use barycentric coordinates to define its 3D position $m_{xyz}$. \newText{(d) shows a more general example. Here, the tiling recipe (TR) has 4 rows and 5 columns. The position of $t.min_{uv}$ determines the reference tile $r$ from the tiling recipe $TR[1,1]$. Based on the size of the triangle, the number of tiles needed to cover the triangle is estimated. In this example, the replication area (\newText{$rep^{R}$}) has 3 rows and 4 columns. The \emph{offset} in the \emph{uv} texture coordinates refers to the vector from $t.min_{uv}$ to the origin of the reference tile $r.min_{uv}$.}} 
   \label{fig:MembranePopulationExample}
\end{figure}
%\begin{figure}[t]
 %   \centering
 %   \includegraphics[width=0.8\linewidth]{figures/012.padding.PNG}
 %   \caption{\rh{Projecting a mesh triangle to the tiling recipe. The position of $t.min_{uv}$ determines the reference tile $TR[1,1]$ from the tiles recipe $TR$. In this example, the tiling recipe is a 2D array with 4 rows and 5 columns. Based on the size of the triangle, the number of tiles ($rep_u$, $rep_v$) needed to cover the triangle is estimated. The \emph{offset} in the \emph{uv} texture coordinates refers to the vector from $t.min_{uv}$ to the origin of the reference tile $origin(TR[1,1])$.}}
 %  \label{fig:Padding}
%\end{figure}
\subsection{Membrane Population}

%TODO:Why do we need a replication area that is larger than the triangle that we want to cover. Can we not map content of the triangle directly. We need to add a discussing the design choice in the beginning and show why it is the right one. 

%\annot{High-level idea}
%To populate the membrane, we cover the mesh of a biological structure with a \rh{2D array from $TR$} based on \rh{mesh} texture parameterization. Each element of this \rh{array} represents a single tile from the tiles recipe $TR$ associated with the mesh. 
Our approach to populate the membrane \rh{overlays} each triangle of the mesh with a \rh{sub-set} of $TR$, \rh{which represents the \emph{Replication Area} ($rep$)}. Based on the texture coordinates we obtain the \rh{replication area} that encloses the triangle \rh{$t$} and then re-projects \rh{that area} with its respective tiles onto the triangle in the 3D space. By
dividing the triangle minimum texture coordinate $t.min_{uv}$ by the size of tile in texture coordinate \newText{${r}.size_{uv}$} we get the two-dimensional index into \rh{tile recipe ($TR_{xy}$)} that represents the starting corner of the replication area \rh{$rep[0,0]$} \rh{(see~\autoref{fig:MembranePopulationExample}. In this example, $TR[1,1]$ represents the starting corner of the replication area.) }

Within this replication area, we populate all molecular instances in parallel. In our case, the \rh{replication area} has always the same dimensions \newText{$(rep^R.dim)$} calculated from the biggest triangle of the mesh and we use it for smaller triangles as well. The \rh{replication area} completely covers the triangle area, but tiles can also lie outside the triangle $t$ area or outside the cell $c$. Overall, for every intersected triangle $t \in c.T$, we run  \newText{$(N^R\times |rep^{R}|)$ } threads \newText{where $N^R$ represent the maximal number of molecular instances that can be found in any rectangular tile $r \in R$}. Threads are associated with a particular molecular instance $m$ that belongs to a certain tile and is stored in a linear buffer. The remaining threads are discontinued. 

\rh{In every population thread, the thread \emph{id} specifies the molecular instance $m$ and the two-dimensional indices within replication area  \newText{$rep^{R}$}. We need to know the two-dimensional index of the tile within $TR$ that corresponds to \newText{$rep^{R}_{ij}$} and from which the populated molecular instance $m$ should be mapped. To achieve that, we can use the minimal \emph{uv} texture coordinate of the triangle to get the two-dimensional index of the starting corner of the replication area within the tile recipe $TR_{xy}$ (see~\autoref{fig:MembranePopulationExample}\newText{-bottom}).%fig:Padding}). %by dividing the triangle minimum $uv$ texture coordinate by ${l}.size_{uv}$. 
 Then, by adding \newText{$(rep^R_i,rep^R_j)$} to the two-dimensional index of the starting corner of the replication area $TR_{xy}$, we get the index of the tile $\newText{r \in R}$}.

\rh{To correctly define the 3D position of the molecular instance $m_{xyz}$ within the tile \newText{$r$}, we need to find its texture coordinate $m_{uv}$ to obtain the barycentric coordinates which will allow us to get $m_{xyz}$ through interpolation.}

The instance texture coordinates $m_{uv}$ can be computed as follows; First, we add $ m_{\mu\gamma}$, which represents its position within the tile, to the tile two-dimensional indices  \newText{$(rep^R_i,rep^R_j)$}. We get the instance position within the replication area in $ {\mu\gamma}$ coordinate. Then, we can map that value to \emph{uv} texture coordinates by multiplying that number by \newText{${r}.size_{uv}$}. Finally, we need to shift that value to $uv$ location that represents the origin of the starting corner of the replication area ( \autoref{fig:MembranePopulationExample}\newText{-bottom}).%fig:Padding})to obtain $m_{uv}$. Now, barycentric coordinates of the instance $m$ within the triangle $t$ can be calculated. Finally, by interpolation, we obtain the 3D position of the instance $m_{xyz}$. }

Afterwards, we crop all the instances that lie outside of the triangle (using barycentric coordinates of the instance) and outside of the cell $c$ \rh{by testing the intersection between the cell's AABB and $m_{xyz}$.} If the position \rh{$m_{xyz}$} passes the criteria, the $atomicCounter$ is increased and a new molecular instance \rh{is recorded in $c.cache[atomicCounter]=m$. The newly populated instance $m$ inherits all the features such as its molecular type, color, and position.}
%Its type is set according to $m.type = m_{ref}.type$ and the position is set to $m.pos = pos$.
The rotation is stored in an analogous way. The only difference with the rotation is, that it \rh{is} adjusted by a rotation representing the rotation around the  \newText{$z$-}axis into the normal vector of the triangle. The \newText{$z$-}axis is used as all the tiles were generated with the default orientation facing the\newText{$z$-}axis. %\rh{The pseudocode of membrane population can be found in~the supplementary material in~\autoref{alg:MembranePopulationNew}.}

\subsection{Solubles Population}
\label{solublespop}
In the previous section, the description of the membrane population was discussed. The next step is to populate internal parts of the biological structures with molecular instances, not the external space on the other side of the boundary. Similar to the membrane population, this approach is based on tiles. However, instead of the rectangular-based tile-set \newText{$R$}, the box-based tile-set $B$ (see \autoref{fig:Patches}) is used. In this case, the population does not rely on the texture coordinates of the mesh. Moreover, there is no Wang tiles approach used for the $B$ tile set. As these are not visible from outside the structure and when immersed inside, it is very unlikely to notice any seams in such a crowded environment. Therefore, the seamless constraint is not applied in the box-tiling case. However, the same Wang tiles concept as previously explained can be extended to 3D and used for $B$ tiles. In our implementation, we generate each box tile $b \in B$ using the same size \rh{${b}.size_{xyz}$}, limited to the cell \rh{$c.size_{xyz}$}.

Firstly, as previously mentioned, \rh{once the cell $c$ is selected to be populated, the set of intersected triangles $c.T$} is computed. Moreover, a list of intersected meshes \rh{$c.H$}, specifying into which meshes the triangles \rh{$c.T$} belong, is created. Afterward, $c.closestTriangle$ (see \autoref{sec:TriangleIntersetion}) is determined.
Every mesh is associated with a box-tile $b$. The box tiles $b \in B$ are tiled inside the cells to fill the internal space of the mesh.
%Analogously to the membrane population, we use the highest number of elements within \newText{a box tile $N^{B}$} together with the number of box tiles that fill the cell $c$,  \newText{(i.e. $|rep^{B}|$)}. The number of threads is calculated as a product \newText{$N^{B} \times |rep^{B}|$}. 
\newText{Analogously to the membrane population, we populate all molecular instances in parallel. The replication area for any cell $c$ has the same dimensions $(rep^{B}.dim)$, representing the number of box tiles needed to fill $c$ completely with box tiles. We run ($N^{B} \times |rep^{B}|$) threads where $N^{B}$ represent the maximal number of molecular instances that can be found in any box tile $b \in B$.}

\rh{Every thread is} associated with a molecular instance $m$, \rh{where all instances} are stored in a linear buffer of the box-tile $b$. For every instance, in a box-tile $m$, its relative 3D position inside the box \rh{$m_{xyz}$} is computed. To calculate the absolute position, we need to calculate the world space position of the starting corner of the box-tile $b$. This position is calculated from the starting corner of the cell $c$, the three indices $x,y,z$ that refer to the respective box-tile $b$ and the world-space size of the box-tile \rh{$b.size_{xyz}$}. Once the 3D position of molecular instance $m_{xyz}$ is calculated, it is tested whether it lies inside the cell $c$. Moreover, if the cell is intersected by triangles \rh{(i.e. $c.T \neq \emptyset$)}, the algorithm decides in which half-space with the respect to \rh{$c.t$} the position \rh{$m_{xyz}$} is. This orientation is determined based on the normal vector of the triangle mesh, whether the normal points toward the \rh{$m_{xyz}$} wrt. the triangle center or away from \rh{$m_{xyz}$}. If \rh{$m_{xyz}$} is outside the biological structure, it is rejected and the computation stops. \rh{Otherwise, the atomicCounter is increased and a new molecular instance is recorded in $c.cache[atomicCounter] = m$. The newly populated instance inherits all the instance $m$ features such as its molecular type, position, and orientation.} For the case when there is no intersection with any mesh \rh{(i.e. $c.T = \emptyset$)}, the closest triangle information $c.closestTriangle$ is used as an indicator, and the triangle normal of the closest triangle is again used to determine whether the entire cell $c$ is inside or outside of the biological structure defined by the mesh. In case it is marked as outside, no population is performed. %\rh{The pseudocode of soluble population can be found in~the supplementary material \autoref{alg:solublePopulation_New}.}

%=======================
\section{Parallel RTX-based Molecular Rendering}
\label{sec:Rendering}
\begin{figure*}[t]
    \centering
   \includegraphics[width=0.9\linewidth]{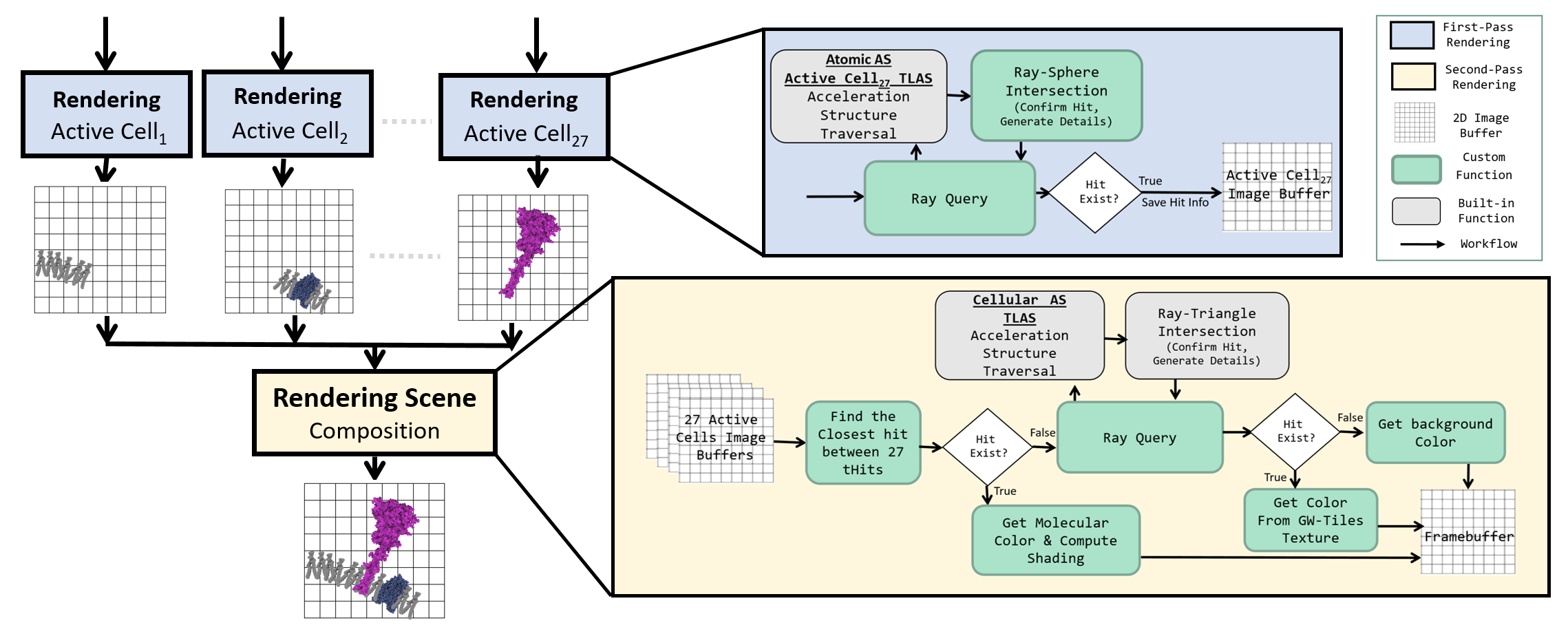}
    \caption{Overview of the parallel rendering pipeline. On the left side, the "sort-last" scheme of parallel rendering is presented which consists of two passes. The first pass is highlighted with blue boxes that trace the Atomistic AS TLASes in parallel while the yellow box represents the second pass that composites the 27 image buffers. On the right side, the corresponding high-level description of each pass.}
   \label{fig:rendering}
\end{figure*}

%\annot{motivation}
The computational complexity of interactive ray tracing grows logarithmically with the complexity of the scene, which challenges the predominant role of rasterization in complex environments~\cite{Wald-2001-State-of-the-Art-in-Interactive-RayTracing,Wald-2001-Interactive-Distributed-RayTracing-Highly-Complex-Models,Wald-2002-OpenRT}. Recent graphics card architectures support hardware accelerated ray tracing. We utilize this new technology to accelerate the rendering of our on-the-fly populated scene.

\subsection{Acceleration Structures}
%\annot{background}
%The acceleration structure {\it (AS)} is a core component of every efficient raytracing algorithm. To accelerate the raytracing in the modern GPUs, this component is implemented in hardware. NVIDIA GPU hardware implementation exposes only two levels of acceleration structure to the user. The {\it bottom-level AS (BLAS)} defines the geometric description of a molecular model (i.e., the position and radius of its atoms), while the {\it top-level AS (TLAS)} consists of instances that reference to one of the BLASes~\cite{RayTracingGems}. Each instance is associated with the transformation matrix, as well as the molecular type of the instance to fetch the corresponding color value. This two-level hierarchy allows us to populate multiple instances of an object while storing its geometry only once in the GPU memory. 

The acceleration structure {\it (AS)} is a core component of every efficient raytracing algorithm. To accelerate the raytracing in the modern GPUs, this component is implemented in hardware. NVIDIA GPU hardware implementation exposes only two levels of acceleration structure to the user. The {\it bottom-level AS (BLAS)} defines the geometric description of a model, while the {\it top-level AS (TLAS)} consists of instances that associated with the transformation matrix, as well as a reference to one of the BLASes~\cite{RayTracingGems}.% This two-level hierarchy allows us to populate multiple instances of an object while storing its geometry only once in the GPU memory. 

\begin{figure}[t]
    \centering
   \includegraphics[width=0.8\linewidth]{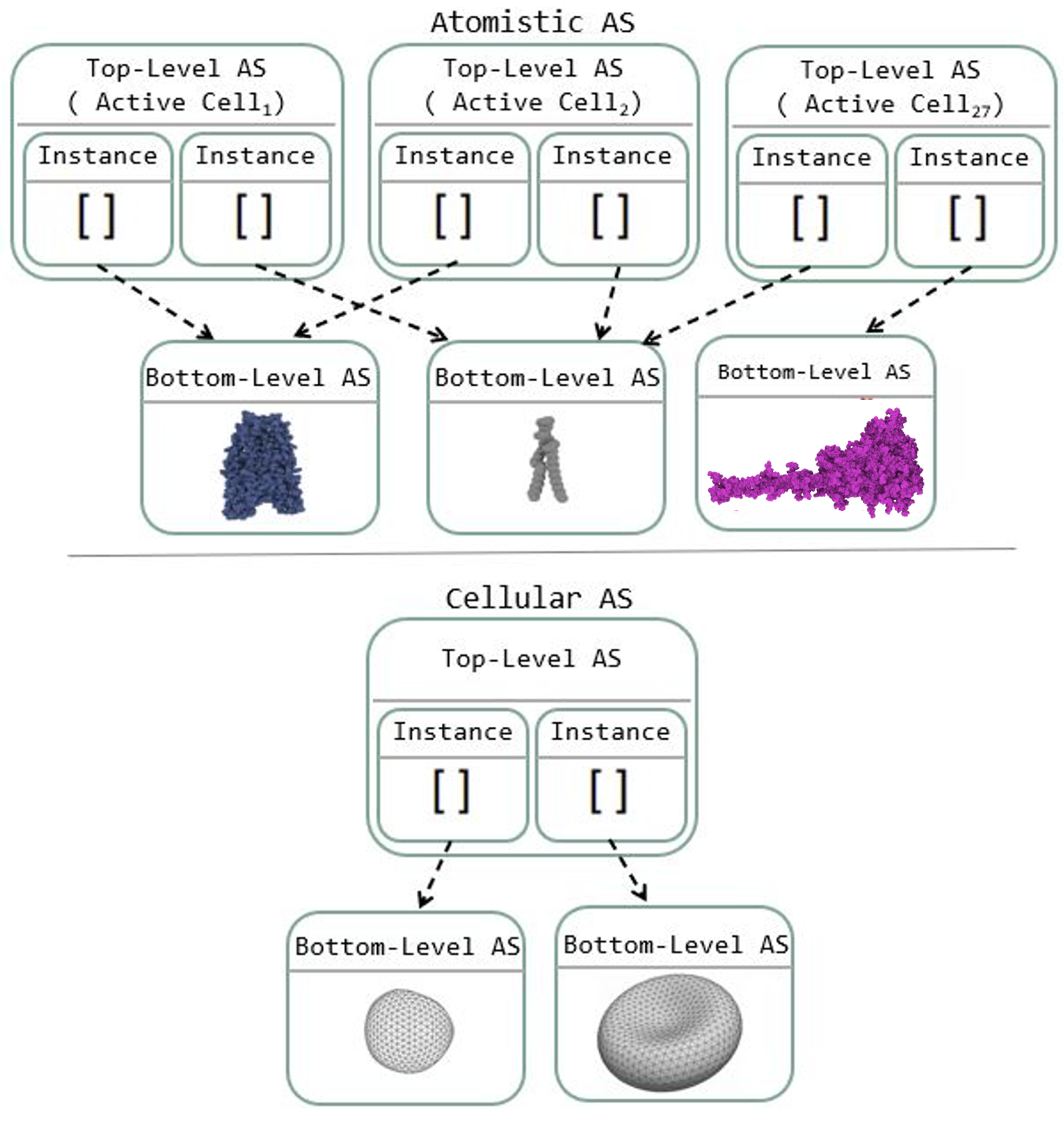}
    \caption{This image shows two types of acceleration structures used by the renderer at the top atomistic AS, which contains the atomistic scale description. The atomistic AS contains several TLASes, a TLAS per active cell. Every TLAS has several instances that point to one of the BLASes. At the bottom is the cellular AS, which contains only one TLAS representing the skeleton of the scene.}
   \label{fig:SceneAS}
\end{figure}

%\annot{two types of AS}
In our rendering algorithm, we are using two representations of acceleration structures: the {\it cellular AS}, which defines the skeleton of the scene and contains all the mesh instances that define the shape, size, and position of the biological structures, and {\it atomistic AS} which contains the atomistic description of the active cells.  In {cellular AS} we create a BLAS for every mesh, while in {atomistic AS} we create a BLAS for every molecular model where we define the position and radius of its atoms, and then we instantiate them within the scene (see \autoref{fig:SceneAS}). 

%\annot{Why atomistic AS has multiple TLASes}
Representing the {atomistic AS} as a single TLAS will lead to rebuilding it from scratch whenever the active cells are changed. RTX acceleration structure allows updating TLAS, which is cheaper than rebuilding it, however, it can be used only for updating the instances information e.g. transformation matrix. If a new instance needs to be added to the scene, the TLAS has to be rebuilt. To avoid that, our {atomistic AS} contains multiple TLASes, one TLAS per active cell. The active cell's TLAS is generated based on the contents of its cache. Once a cell becomes active, its TLAS will be built and will not change until that cell becomes inactive.

%\annot{ray intersection test}
For each TLAS, hardware acceleration requires providing the type of ray intersection test needed by the traversal program. The selection should be based on the BLAS lowest geometric representation. In the {\it cellular AS}, the meshes are defined as triangles, therefore we use the hardware triangle-ray intersection test that is built-in in the GPU. In the {\it atomistic AS} the molecular models are defined through its atoms/spheres, therefore we use a custom-implemented sphere-ray intersection test.

\subsection{Parallel Rendering}
%\annot{High-level idea}
The scene is rendered as a combination of atomistic and cellular rendering. Details closer to the camera are rendered at the atomistic resolution, while objects further away from the camera are rendered as textures. 
%This provides us with a higher-performance texture. % as the texture has a much smaller memory footprint than the geometric representation. 
To prevent the sudden popping of atomistic structures, we implement a smooth transition from texture details into atomistic details and vice-versa using alpha blending. 
%\annot{How we parallelize the rendering}
In the rendering of the molecular details, the scene is rendered in two-pass rendering. In the first pass, the active cells are rendered separately into their respective frame buffer objects (FBOs). This pass takes the advantage of having multiple TLASes in the {atomistic AS} to parallelize their rendering. We use the {\it sort-last} parallel rendering scheme~\cite{Molnar-1994-A-sorting-classification-of-parallel-rendering} that can render the  {atomistic AS} TLASes in parallel, which results in a very high data rate as the rendering operates independently. But instead of parallelizing the rendering through multiple GPUs, we use one GPU that uses compute shader and Nvidia's \qcrFont{GLSL\_EXT\_ray\_query} extension to parallelize the rendering tasks between threads. The ray query extension allows us to invoke ray tracing queries through the compute shader. This extension is an alternative to the ray tracing pipeline, but no separate dynamic shader or shader binding table is needed~\cite{RayQueries}.{ To our knowledge, no technical literature has reported using \qcrFont{GLSL\_EXT\_ray\_query} extension for parallel rendering so far.}
%\annot{Two render passes}
In the first rendering pass, the atomistic AS TLASes are traced in parallel, a thread per pixel per active cell. In each thread, once the closest hit is found, its information (\eg~depth, {\it instance\_id}, {\it atom\_id}) is stored on the full screen image buffer of the thread's active cell. Otherwise, the value ($-1$) is stored, which means the ray did not hit any nanoscale structure (see \autoref{fig:rendering}).

In the second pass, the resulting FBOs are composited, based on the depth values to form the final rendered image. If there is at least one hit found in the 27 images, the {\it instance\_id} and {\it atom\_id} of the closest hit among them is used to get the molecular color. In addition, the shading is computed using {\it Phong illumination model} as well as {\it ray-traced ambient occlusion}. If the is no hit (no nanoscale structure information is found), the cellular structure information is provided using an image-based approach.

\subsection{Image-based Tiling}
%\annot{motivation}
Image-based impostors are usually used to avoid rendering objects that are far away from the viewpoint by replacing the geometry of these objects with a painted texture~\cite{aliaga-1999-Automatic-image-placement-to-provide-guaranteed-frame-rate,Aliaga-1999-MMR,schaufler-1996-Three-Dimensional-Image-Cache}. %We are using the seamless texture map that has been generated from the GW-tiles to assign a color to those far structures. %This texture with {cellular AS} is used to assign a color to that far structures.
%\annot{Seamless GW-tile texture}
In the tile preparation phase (see \autoref{sec:Tiles-Preparation}), GW-tiles have been created. Moreover, the corresponding texture map was synthesized. The key idea is to use both of these levels of detail representations while rendering the scene. When the camera is closer to a biological structure, the GW-tile is used. Once the camera zooms out, which causes the atomistic detail to disappear, the corresponding part of the texture map is rendered in the very same place. In  \autoref{sec:Tiles-Preparation} also the description of tile recipe $TR$ was presented. The tile recipe forms a virtual map of tiles that covers the whole $uv$-texture space associated with the mesh.

The texture map is sampled while rendering a cellular mesh using the following approach. Using the texture coordinate of a fragment \newText{$f_{uv}$}, we first determine which tile it lands at, based on the position \newText{$f_{uv}$} within the tiles recipe. Moreover, the relative \rh{fragment position\newText{$f_{\mu\gamma}$}} inside this tile is computed. Based on the \rh{tile \emph{id}},  \rh{the tile's} respective starting position \newText{$tileTex.min_{uv}$} in \rh{the} texture map is obtained. The resulting color is fetched for the position \newText{($tileTex.min_{uv} + \newText{f_{\mu\gamma}}$)} from the packed texture map.
%\rh{The pseudocode of sampling from the packed texture map can be found in~ \autoref{alg:sample-packed-texture-New} in the supplementary material.}
Besides the diffuse color, the texture map consists of normal and ambient occlusion buffers which are used to add geometric detail to the shaded surface.

%To map the texture coordinate of the mesh to the GW-tile texture, we use the following algorithm. For remark, we work with 16 base texture tiles. These 16 tiles are uniformly distributed in the GW-texture in the way, where there is one additional extra tile that fits the Wang-Tile pattern. Together, a texture consisting of 9x9 tiles (see \ref{fig:Textures}) is created. 

%\annot{smooth transition from geometry to texture}
To avoid the sharp transition from geometry to image-based representation we apply alpha blending on the instances that are located in the far border of the neighboring cells which combines the geometry atomistic color with image-based cellular color. \rh{Due to high-frequency components of the image-based texture, texturing far meshes results in aliasing artifacts. To avoid that, we use two-level hierarchy of colors for the meshes where the first is the packed Wang tiles texture while the other is a solid color and blend them based on the world space distance to the camera.}

%=======================
%\section{Technical Implementation}
%\input{sections/TechnicalImplementation}
%=======================
\section{Results}
%\annot{Our contribution}
\begin{figure*}[t]
    \centering
    \includegraphics[width=1.0\linewidth]{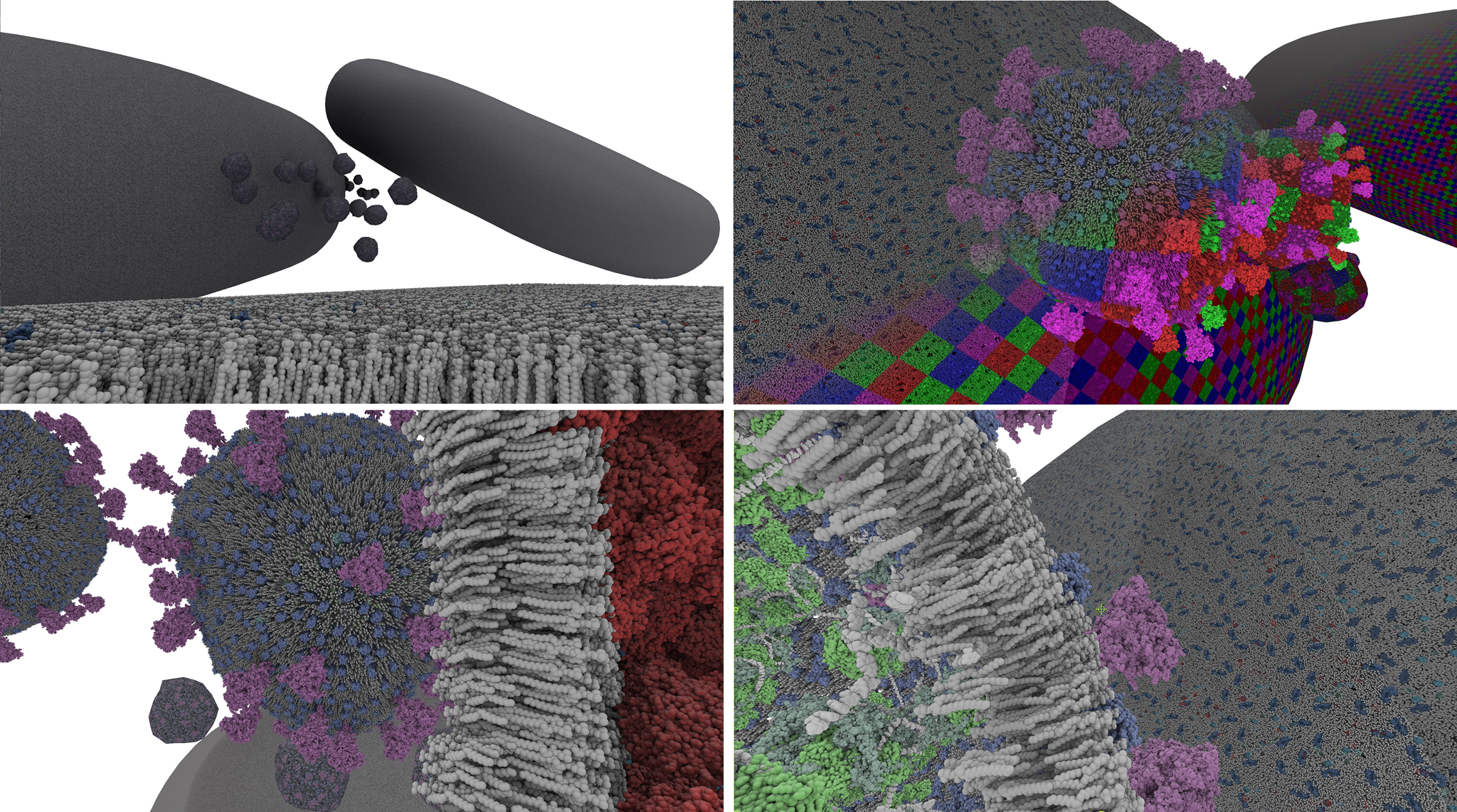}
    \caption{Example of the resulting scene rendered using our RTX pipeline. Top-Left: partially populated RBC membrane with an overview of the scene. Top-Right: Several populated SARS-CoV-2 particles \rh{partially overlayed by continuous tiling}. Bottom-Left: View from the populated RBC membrane towards partially populated and not populated SARS-CoV-2 particles. Bottom-Right: View inside SARS-CoV-2 particle. \rh{Higher resolution images can be found in Appendix~\ref{appendix:ResultImagesInHighResolution}.}
    }
   \label{fig:Result}
\end{figure*}

\begin{table*}[t]%  R3.8
\centering
\caption{\rh{Biological models used in the final model scene. The table shows the axis-aligned bounding box (AABB) in {\AA}ngstrom for the mesh of each model, and shows the number of molecules (\#m) and atoms (\#a) for the full model and its rectangle and box tiles. }}
{%\color{purple}
\scalebox{0.8}{
\begin{tabular}[t]{l|c|ccc|ccc|cc}
\hline
 Biological model & AABB & \multicolumn{3}{c|}{rectangle-tile}&\multicolumn{3}{c|}{box-tile}& \multicolumn{2}{c}{full model}\\
 & (\AA) &size (\AA) &\#m  & \#a &size (\AA) &\#m  & \#a&\#m & \#a \\
\hline
SARS-CoV-2&$1,187 \times 1,166 \times 1,162$&$500 \times 500$&8.2K&886K&$1,158\times1,107\times 1,133$&53K&4.5M& {$135$}K & $\sim 24$M\\
RBC& $64,360 \times 15,110 \times 75,410$ 
&$500 \times 500$&5K&335K&$1,000\times1,000\times 1,000$&$5$K & 21M& \newText{$518$} M& $\sim 1.2$T\\
\hline
\end{tabular}}}
\label{table:Dataset}
\end{table*}

% SARS has 136,522 instances (23.7209 million atoms)
%RBC has 518.367 million instances ( 1.02286 trillion atoms ) 

Our novel construction approach is capable of instantly generating and visualizing biological worlds of cellular mesoscale.
We demonstrate the scalability of Nanomatrix on a scene containing a red blood cell (RBC) and a SARS-COV-2 virion, as shown in~\autoref{fig:teaser}.
  
%Each cell of the red blood cell contains around ?? molecules with ?? atoms.  Each cell of a SARS-COV-2 viruses contains around ?? molecules with ?? atoms. Since we using 27 active cells we are keeping around approx. ?? molecular instances on average in memory.  If fully generated, the whole scene would contain ?? molecules with ?? atoms.
% RESULTS
% MEMBRANE
% the area of one tile is 500x500=250000 \AA^2
% the area of RBC mesh with scaling 40000 (range of vertices [-1,-] is 9.54061e+09 \AA^2  => x area of tile
% 12,784,270,000 atoms => atoms for membrane
% 
% SOLUBLE
% bounding box size of RBC (scaling 40000): 64364.6 x 15116.2 x 75412.6  = 73,334,686,636,000 \AA^3
% one brick is 1000 x 1000 x 1000 = 1 000 000 000 \AA^3
% one brick has 1000 molecular instances, with approx. 4 400 000 atoms
% 322,672,621,198 => atoms
% total: 335,456,891,198

% hemoglobin approx. 4380 atoms
% $9.54061e+09 \AA^2$ / 250,000AA2 = 38,162
Instantly generating and rendering such a large scene in atomistic detail was not possible with previous methods. The bounding box size of RBC mesh is $64,360 \times 15,110 \times 75,410= 73,334,686,636,000\AA^3$. The generated box-tile size is $1,000 \times 1,000 \times 1,000 = 1,000,000,000 \AA^3$. Each box tile has approx. $5,000$ molecular instances (hemoglobins of approx $4,380$ of atoms), with approx. $21,000,000$ atoms. % what is the "emptiness ratio of a donut inside the box?" 30%?
Additionally, the surface area of the RBC mesh is $9.54061e+09 \AA^2$. The area of its membrane GW-tile is $500 \times 500 = 250,000 \AA^2$. Each membrane GW-tile has approx. $5,000$ molecular instances with approx $335,000$ atoms. 
Once the RBC would be fully generated, the entire scene would contain approx. \newText{$518$} millions of molecules with $1.2$ trillion of atoms. %Additionally, the surface area of the RBC mesh is $9.54061e+09 \AA^2$. The area of its membrane GW-tile is $500 \times 500 = 250,000 \AA^2$. Each membrane GW-tile has approx. $5,000$ molecular instances with approx $335,000$ atoms. Therefore, the fully populated surface would consist of approx. $175$ million of molecular instances with approx. $13$ billion of atoms. 
One SARS-CoV-2 virion consists of approx. $135$ thousand molecular instances with approx. $24$ million of atoms \rh{(see~\autoref{table:Dataset} for more details)}. In our model scene, there are four RBCs and twenty SARS-CoV-2 particles which leads to approx. $5$ trillion of atoms in total. However, the algorithm is scalable enough to work with any number of non-overlapping models that can be fitted into the bounding box of the scene.
The implementation of the approach was realized using the Vulkan API~\cite{VULKAN} and NVIDIA's nvpro-samples framework~\cite{nvpro-samples}.
The performance was measured using a NVIDIA GeForce RTX \rh{4090 graphics card with 24 GB memory}. 
 In our experiment, we create a grid of dimension $200\times200\times200$ with cell size $2000$. Each cell cache buffer has room for \rh{$500,000$} molecular instances. These are programmatically adjustable, they can be set to meet the requirements of any dedicated GPU. 
 %=>TODO Performance measurements for the construction and rendering?
The construction algorithm is able to populate a cell with membrane molecules in approx. 6 ms and with soluble molecules in approx. 3 ms.
%\annot{MesoCraft vs. RTX rendering}
We implemented the approach in two environments. The first environment, where the rendering is built on top of the Marion library, is developed using C++ and OpenGL 4.6 graphics API~\cite{Mindek-2018-marion}. The second environment is built using the Vulkan graphics API with RTX functionality. \rh{\autoref{fig:Result} shows our model scene rendered using \rh{RTX-based renderer}.} Whereas the construction algorithm has approx. the same performance (as in both cases it relies on the compute shader pipeline and not on the graphics pipeline), the rendering differs significantly. We are able to achieve on average factor 2.5 speedup using the  \rh{RTX-based} framework. On NVIDIA GeForce \rh{4090} in full HD resolution with similar settings of the scalable construction algorithm, Marion rendering runs at approx. \newText{55~FPS} (with drops down to \newText{35~FPS}). The \rh{RTX-based} framework runs at \rh{approx. 150~FPS (with drops to 110~FPS)} with a single ray per pixel and \rh{110~FPS (with drops to 80~FPS)} with 10 rays per pixel. This speedup is sufficient enough to provide us with the future possibility of integrating a VR or AR interface, where a high framerate is the key factor for a satisfactory user experience. The performance of Nanomatrix can be further boosted by submitting the rendering and construction workloads into two different \emph{Queue Families} in the GPU which allows the RT Core and compute workloads to be processed concurrently. This is a new feature that is supported by recent NVIDIA’s graphics card architectures~\cite{NVIDIA-2021-Ampere-GPU-Architecture-whitepaper}.

\begin{figure}[t]
    \centering
    \includegraphics[width=1.0\linewidth]{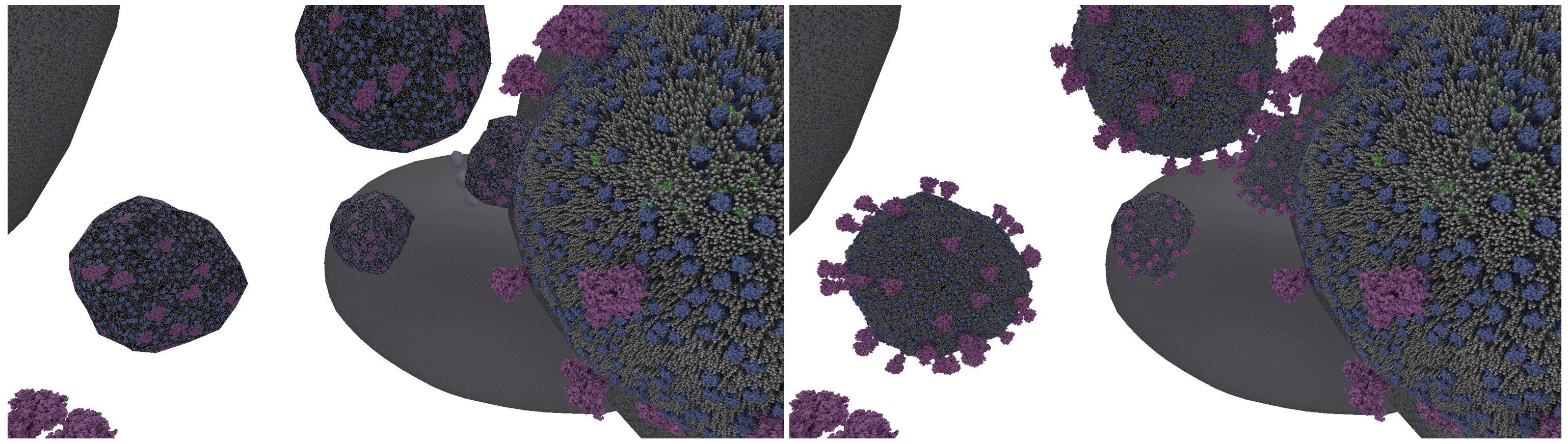}
    \caption{\rh{Result of two different activation window layouts. On the left is the result of activating the closest neighbor cells while on the right the closest neighbor cells that intersect the frustum are activated.}}
   \label{fig:cellSelectionMethod}
\end{figure}

\section{Discussion}
%\annot{Object-Space AO - inaccurate shading }
We implement Object-Space AO (OSAO) to convey the shape and the depth of molecules by tracing random AO-ray against the TLAS of the active cell to which the hit primitive belongs. %\rh{(see~\autoref{fig:AO} in the supplementary material)}. 
The AO algorithm is implemented as described in NVIDIA Vulkan API tutorials~\cite{NVIDIA-Ambient-Occlusion}. Due to our griding approach, the AO is inaccurate on the primitives that are located on the borders of the active cells. To estimate the shading value correctly, the TLAS of the adjacent cell would need to be traced as well. This is one of the drawbacks of the sort-last scheme of parallel rendering~\cite{Dietrich-2007-Massive-Model-Rendering-Techniques}. To overcome this issue, a test has been added in the AO computation, if the hit atom is located on the cell border, then the AO-ray will traverse the AS of all active cells that intersect the hit atom.

\rh{Selecting which neighboring cells should be activated could
be done based only on the camera position or based on both the camera position and direction by activating the closest neighbor that
intersects the view frustum. \autoref{fig:cellSelectionMethod} shows side-by-side comparison of the methods result. While the frustum-based selection method enriches the scene with geometrical information, the camera position-based method could be more suitable in immersive environments, such as VR, where camera position/direction is continuously changing and using the frustum-based selection will cause more populating/destroying cells.}

The instance count at the RTX acceleration structure must not exceed a specific limit which varies based on the graphics card.  Our \rh{RTX 4090} graphics card supports a maximum 16M instances. However, the scalability of Nanomatrix allows overcoming that limit by rendering multiple acceleration structures in parallel. {The selected design for the atomistic acceleration structure (see~\autoref{fig:SceneAS}, top) has several additional advantages. In this acceleration structure, each BLAS defines the geometric description of a molecular model, while the active cells TLASes consist of instances associated with the transformation matrix, as well as a reference to one of the BLASes. This two-level hierarchy allows us to populate multiple instances of a molecule in several TLASes while storing its geometry only once in the GPU memory. In addition, it enables us to construct/destroy cells independently, which supports parallel cell populations. Another advantage of this acceleration structure design, it fits nicely with the sort-last parallel rendering scheme that accelerates the performance as  the rendering operates on the TLASes independently. We have experimented with various ways for how the RTX’s acceleration structure can be defined for Nanomatrix. We finally converge on an optimal solution with the proposed construction algorithm that leads to a high rendering performance while maintaining the memory requirement low and updatable.}

Our method is scalable, its parameters can be adjusted based on the available computational resources. As the size increases, more geometrical information will be presented which enriches the scene with detailed information. However, it will increase the computational complexity and the memory footprint. Our method allocates a part of the dedicated GPU memory for the cells' cache buffers. Clearly, increasing the cache buffer size will increase the allocated portion of the memory. On the other hand, increasing the size of the activation window will increase the number of the cells' cache buffers. In addition, the rendering overhead increases with the size of the activation window, because it requires more rendering threads in the first-pass rendering and more images to be composited in the second-pass rendering. %\rh{There is extensive performance evaluation available in supplementary material, Appendix~\ref{appendix:PerformanceEvaluation}.} 
\newText{Based on our experimentation, we have set up the cells size to $2,000\AA$, and we activate each time only 27 of them surrounding the camera. Subsequently, we use geometric representation for close objects and change to image-based representation if the distance to the object is more than $3,000\AA$ (assuming the camera is in the middle of the central cell). These have been hand-crafted and serve the test on our system, NVIDIA GeForce RTX 4090, which was able to run the framework at approx. 150~FPS (with drops to 110~FPS) in full HD resolution. On another system, there could be another setup. It would be natural to design an adaptive approach that controls the selection of visual representation such that a desired rendering frame rate is secured. However, such automation is out of scope of this paper and would be an interesting future work investigation.% The actual securing that you define the number of frames per second, and we start with a more general setting and then through a feedback loop you evaluate whether you reach that FPS and if you do not then you lower your visual representation until this is met then the system can set theses parameters based on this performance analysis. Such an approach would be interesting for future work investigation.
}

\newText{Our construction algorithm is meant for explanatory visualization of extremely large cellular mesoscale scenes that can be explored down to atomistic detail. The tiling strategy that we employ may be criticized for repetitiveness and associated plausibility of the resulting model. We want to emphasize that the explanatory visualization scenario, like an interactive show in a science center for example, allows for certain flexibility in terms of structural accuracy of the biological system that might or might not be acceptable within scientific discovery workflows. Presented biological scenes are accurate with respect to the concentration and the density of the structural composition. While there are no repetitions in depicted biological systems, our construction creates a seamless repetition of structures. As said, this is tolerable for broad audience use-case as it is the case in many graphics applications where Wang-tiling is frequently used.} 

%===========================================
\subsection{\newText{Comparison to Existing Work}} 
Nanomatrix is a view-guided construction system where the construction algorithm is paired with visualization that makes it possible to visualize multiple RBC instances in one scene (four in our example case). No in-core algorithm can render a single RBC in atomistic detail because it doesn't fit into the current sizes of GPU memory. A single RBC contains approx. 1.2 trillion atoms. Saving just the position (3$\times$ \texttt{float}) and type (1$\times$ \texttt{byte}) for an atom requires (13$\times$ \texttt{bytes}). This means a single RBC requires 15.6 trillion bytes (approx. 1.5TB). 

For testing purposes, we exported a single cell of RBC of size $2000 \times 2000 \times 2000 \AA$, which contains approx. 49.5 million atoms (shown in supplementary material). Firstly, we used off-the-shelf visualization tools Avizo~\cite{Avizo} and ParaView~\cite{paraview}. As the strength of these tools is in user-defined custom visualization of generic data, they are not optimized for our dataset and they crashed. %IV: on which hardware?

Then, we tested this model in the open-source molecular data renderers available in MegaMol. The RTXPkD~\cite{RTXpkd} renderer which uses a kD tree adapted to particles with hardware raytracing and the multilevel culling variant by Grottel et al.~\cite{OSMegamol}, which is optimized for rendering large data. %Both tests ran on GeForce RTX 4080 GPU (16 GB) in full HD resolution. 
The results are shown in supplementary material. %~\autoref{fig:nanomatrixVSMegamol}. 
The left image shows the model in RTXPkD renderer which ran with 83 FPS while the middle image shows the result from Grottel et al.~\cite{OSMegamol} renderer which ran with 177 FPS. However, testing 27 active cells in MegaMol was not possible because of the GPU memory limitations. Visualizing 27 copies of this model requires at least 17,5 GB of GPU memory, which does not leave much room for framebuffer, and other data structures necessary for the renderer.   

Several molecular renderers, such as Marion~\cite{Mindek-2018-marion}, exploit instancing for visualizing large molecular data as the biological models often consist of a large number of recurring molecules. The instance-based scheme reduces the amount of required storage by defining molecules only once and instantiating them many times within the scene. However, these renderers still cannot visualize a single RBC. Saving only the position, rotation, and type of an instance require 29 \texttt{bytes}. A single RBC contains approx. 518 million of instances of molecular structures. Therefore, it requires in the most optimal case approx. 15 GB of GPU memory. However, the implementation requires more memory and therefore it is not possible to fit the model fully in. %IV: for which VRAM size?
For ray-tracing, additional BVH data further increases the memory overhead. BVH requires almost 3.5 times the raw data size~\cite{RTXpkd}, as a bounding box needs to be stored for each node in the hierarchy. Even with the instance-based scheme, none of the current techniques are at the moment able to render one RBC in a GPU with 24 GB memory. %Or render more than 600 million instances in GPU with 48GB memory. 

The main idea of our approach to overcome the limitations (memory, amount of instances) is to \emph{"compute instead of store"}. The whole enormous scene is never completely stored in the memory, only a fraction of it, which is close to the viewer. The only approach that follows the same gist is the Instant Construction~\cite{Tobias-2018-Instant-Construction} approach, however, this algorithm is not scalable and cannot construct/visualize worlds that are larger than what can be packed into the GPU memory. They don't have any scheme that would be able to partition the scene and show it on demand. Our approach can be considered as a scalable version that is built on top of this non-scalable prior work.  

All the in-core systems are limited by GPU memory. Popular tools such as ParaView, VisIt, and VMD support distributed rendering when data is too large to fit in a single GPU. Investigating this goes far beyond the scope of this work. These tools also integrate Intel OSPRay~\cite{OSPRay} into their systems. OSPRay~\cite{OSPRay} is a scalable, CPU-based ray-tracing library for interactive applications. It supports instancing and is designed for visualizing large data as long as it fits the available CPU memory. Recently, Intel introduced OSPRay Studio~\cite{OSPRayStudio2021} which is an open-source and interactive visualization that leverages Intel OSPRay~\cite{OSPRay} as its core rendering engine. We implemented a plugin into OSPRay Studio for importing our molecular models to investigate the possibility of rendering our models using OSPRay. Performance was measured on Intel Xeon Gold 6242 2.80 GHz at HD resolution. %\autoref{fig:SARS_OsprayVsNanomatrix} shows the visualization of SARS-CoV-2 in OSPRay Studio at the top compared to our approach at the bottom. 

We exported the content of eight active cells of size $2000 \AA$ of RBC model which consists of 1,322,381 instances (1,069.97 million atoms). OSPRay took approx. 12.5 hours to build its internal data structure and to prepare the scene. It ran with 3.21 FPS. Based on Windows Task Manager, OSPRay Studio application was utilizing 18.398 GB of RAM and 93\% of CPU. 
 Our approach is able to construct the same model in approx. 67 milliseconds and render it with 130 FPS. We exported the content of 27 active cells of size $2000 \AA$ of RBC model. %(see~\autoref{fig:OSPRay_RBC_27cells}).  
 This model consists of 3,296,281 instances (1,748.9 million atoms).
OSPRay took approx. 3 days to finally render it with 1.14 FPS. The application was utilizing 45.537 GB of RAM and  93\% of CPU. Nanomatrix is able to construct the same model in approx. 97 milliseconds and render it with 100-85 FPS.

By integrating the construction into the rendering, we overcome the limitations of streaming pre-generated data to memory. It is essential to emphasize that the focus of this paper lies not only in rendering but especially on scalable construction, coupled with rendering. Our main contribution centers around the scalability of our approach to render four RBCs and possibly even many more. Notably, considering the microscopic scale of RBC is only 8~{\textmu}m  and there are many larger biological cells that need to be visualized.

%=============================================
\subsection{\rh{Limitation}}
\begin{figure}[t]
    \centering
    %\includegraphics[width=1.0\linewidth]{figures/008.tiling-limitation.png}
    %\caption{\rh{Illustration of the relation of texturing quality (left column) to the resulting populated model (right column). Top row: the parts of the mesh with undistorted texture mapping. Middle row: every triangle of the mesh is textured independently. Bottom row: distorted texturing of an arbitrary mesh. The middle column illustrates TR application onto the mesh.}}
      \includegraphics[width=1.0\linewidth]{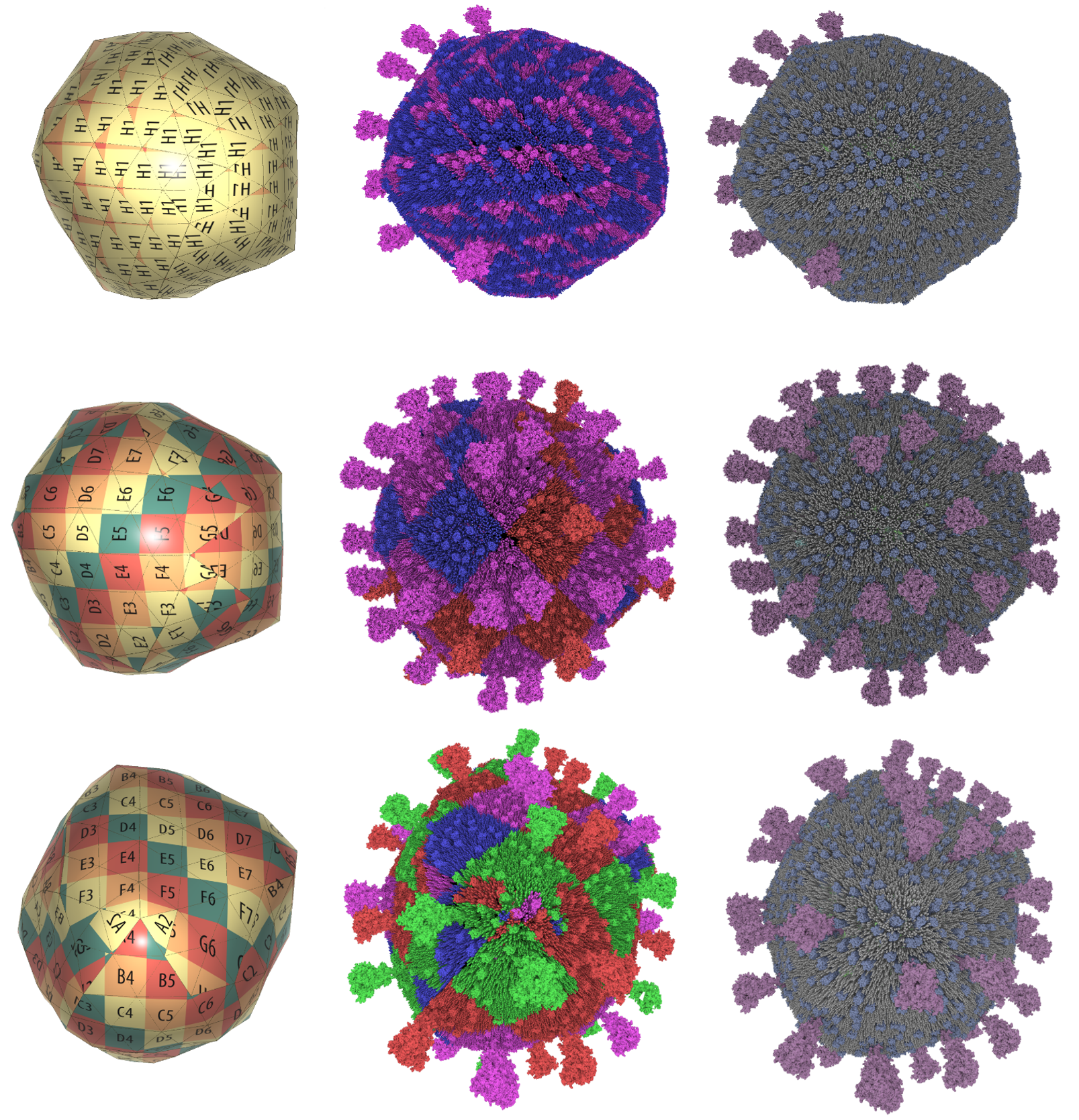}
    \caption{\rh{Illustration of the relation of texturing quality (left column) to the resulting populated model (right column). Top row: every triangle of the mesh is textured independently. Middle row: the parts of the mesh with undistorted texture mapping. Bottom row: distorted texturing of an arbitrary mesh. The middle column illustrates TR application onto the mesh.}}
   \label{fig:tiling-limitation}
\end{figure}

{Unlike the existing approaches, our tiling approach is not restricted to mapping a tile into a single face of the same size.}
\rh{It can be applied on arbitrary meshes with arbitrarily-sized triangles. The quality of mesh texturing plays the key role. In the areas of meshes with continuous texturing, our method populates elements with no visible seams. There still may occur seams mainly around the texture wrapping (see \autoref{fig:tiling-limitation}).}

%=======================
\section{Conclusions}
This work presents a scalable approach for exploring cellular mesoscale down to their atomistic resolution. We introduce a view-guided construction algorithm based on the Wang tile concept. \rh{ Our implementation uses Wang tile concept for construction of membrane structures of biological entities. Construction of internal parts of biological entites does not integrate the Wang-tile concept, but the extension is straightforward.} The overall performance is interactive even for hundreds of billion of atoms.
Currently, the box-tiling is not aperiodic as the seams are not critical. However, in case we include the representation of linear strands of genetic macromolecules or other fibers, the continuity will become a clear requirement. As the approach which is used for rectangular tiling can be extended for 3D, we can include the fibrous structures as parts of the GW-tiles. If such approach would significantly increase the GPU memory requirements, an on-the-fly fiber population within active cells can be integrated building on the top of Klein et al. parallel fiber generation approach~\cite{Klein-2019-Parallel-Generation-and-Visualization-of-Bacterial-Genome-Structures}. The new active cells will need to populate the fiber from its last position from previous active cells.

\rh{Occlusion culling is a popular strategy for determining which scene to load or which scene to compute or which scene element to render. CellView~\cite{LeMuzic-2015-cellVIEW} utilizes the hierarchical z-buffer for accelerating the rendering process however the largest object is a  macro molecule. It would be interesting to investigate an occlusion culling technique that is designed for a higher range of spatial scales up to the entire cell and potentially beyond. Such an approach would be interesting for future work investigation.}

\newText{Nanomatrix generates geometric representations of structures that are close to the camera and uses image texture representations for structures that are far away from the camera that their atomistic detail is no longer perceivable. The seam between the two representations exists because they are conceptually different. We apply alpha blending to reduce the seam, however, there are several ways that could be used to eliminate or reduce this problem. One way is by interpolating the appearance from one representation to another and smoothening it. for example, we can extend the validity of the geometry and draw for every pixel both texture and geometry at the same position and create a linear transition between them. Another way is by dithering the border, so instead of using linear interpolation, we make a decision with probability whether the texture representation will be used for that pixel or the geometry and by that the boundary would be totally smeared out. Another more expensive approach is increasing the perceived geometric complexity of image tiles by using displacement mapping~\cite{Thonat-2021-Tessellation-Free-Displacement-Mapping-for-Ray-Tracing}. In this case, for each mesh face, an AABB represents the displaced surface needs to be stored as BLAS. Then, during the rendering, once the ray hits the BLAS, the algorithm should traverse the corresponding displaced surface through a custom intersection shader to compute the intersection. All these are valid approaches, but finding the optimal solution in terms of both visual quality and performance needs more in-depth investigation which could be the subject of a future work.}

\section*{Acknowledgment}
%====== Acknowledgment =========
The research was supported by the King Abdullah University of Science and Technology (BAS/1/1680-01-01). We thank nanographics.at for providing the Marion library and Guido Reina for helping with testing MegaMol renderers.

\bibliographystyle{unsrt}
\bibliography{main}

\begin{thebibliography}{10}

\bibitem{GOODSELL-2020-Art-and-Science-of-the-Cellular-Mesoscale}
David~S. Goodsell, Arthur~J. Olson, and Stefano Forli.
\newblock Art and science of the cellular mesoscale.
\newblock {\em Trends in Biochemical Sciences}, 45(6):472--483, 2020.

\bibitem{Tobias-2018-Instant-Construction}
Tobias Klein, Ludovic Autin, Barbora Kozlíková, David~S. Goodsell, Arthur
  Olson, M.~Eduard Gröller, and Ivan Viola.
\newblock {Instant Construction and Visualization of Crowded Biological
  Environments}.
\newblock {\em IEEE Transactions on Visualization and Computer Graphics},
  24(1):862--872, 2018.

\bibitem{cellPACK}
Graham~T. Johnson, Ludovic Autin, Mostafa Al-alusi, David~S. Goodsell,
  Michel~F. Sanner, and Arthur~J. Olson.
\newblock cellpack: a virtual mesoscope to model and visualize structural
  systems biology.
\newblock {\em Nature Methods}, 12(1):85--91, 01 2015.
\newblock Copyright - Copyright Nature Publishing Group Jan 2015; Document
  feature - ; Last updated - 2018-11-01.

\bibitem{Varadhan-2002-Out-of-core-rendering-of-massive-geometric-environments}
G.~Varadhan and D.~Manocha.
\newblock Out-of-core rendering of massive geometric environments.
\newblock In {\em IEEE Visualization, 2002. VIS 2002.}, pages 69--76, 2002.

\bibitem{Wald-2019-RTX-Beyond-RayTracing}
Ingo Wald, Will Usher, Nate Morrical, Laura Lediaev, and Valerio Pascucci.
\newblock {RTX} {Beyond} {Ray} {Tracing:} {Exploring} the {Use} of {Hardware}
  {Ray} {Tracing} {Cores} for {Tet}-{Mesh} {Point} {Location}.
\newblock In {\em High-Performance Graphics - Short Papers}, 2019.

\bibitem{merrell2010model}
Paul Merrell and Dinesh Manocha.
\newblock Model synthesis: A general procedural modeling algorithm.
\newblock {\em IEEE transactions on visualization and computer graphics},
  17(6):715--728, 2010.

\bibitem{wonka2003instant}
Peter Wonka, Michael Wimmer, Fran{\c{c}}ois Sillion, and William Ribarsky.
\newblock Instant architecture.
\newblock {\em ACM Transactions on Graphics (TOG)}, 22(3):669--677, 2003.

\bibitem{greuter2003real}
Stefan Greuter, Jeremy Parker, Nigel Stewart, and Geoff Leach.
\newblock Real-time procedural generation of 'pseudo infinite' cities.
\newblock In {\em Proceedings of the 1st international conference on Computer
  graphics and interactive techniques in Australasia and South East Asia},
  pages 87--ff, 2003.

\bibitem{decaudin2004rendering}
Philippe Decaudin and Fabrice Neyret.
\newblock Rendering forest scenes in real-time.
\newblock In {\em EGSR04: 15th Eurographics Symposium on Rendering}, pages
  93--102. Eurographics Association, 2004.

\bibitem{boechat2016representing}
Pedro Boechat, Mark Dokter, Michael Kenzel, Hans-Peter Seidel, Dieter
  Schmalstieg, and Markus Steinberger.
\newblock Representing and scheduling procedural generation using operator
  graphs.
\newblock {\em ACM Transactions on Graphics (TOG)}, 35(6):1--12, 2016.

\bibitem{Steinberger2014}
Markus Steinberger, Michael Kenzel, Bernhard Kainz, Peter Wonka, and Dieter
  Schmalstieg.
\newblock On-the-fly generation and rendering of infinite cities on the {GPU}.
\newblock {\em Computer Graphics Forum}, 33(2):105--114, 2014.

\bibitem{Gardner-2018-CellPAINT}
Adam Gardner, Ludovic Autin, Brett Barbaro, Arthur~J. Olson, and David~S.
  Goodsell.
\newblock {CellPAINT: Interactive Illustration of Dynamic Mesoscale Cellular
  Environments}.
\newblock {\em IEEE Computer Graphics and Applications}, 38(6):51--66, 2018.

\bibitem{Mesocraft}
Ngan Nguyen, Ondřej Strnad, Tobias Klein, Deng Luo, Ruwayda Alharbi, Peter
  Wonka, Martina Maritan, Peter Mindek, Ludovic Autin, David~S. Goodsell, and
  Ivan Viola.
\newblock Modeling in the time of covid-19: Statistical and rule-based
  mesoscale models.
\newblock {\em IEEE Transactions on Visualization and Computer Graphics},
  27(2):722--732, 2021.

\bibitem{Klein-2019-Parallel-Generation-and-Visualization-of-Bacterial-Genome-Structures}
Tobias Klein, Peter Mindek, Ludovic Autin, David~S. Goodsell, Arthur~J. Olson,
  Eduard Gr{\"o}ller, and Ivan Viola.
\newblock Parallel generation and visualization of bacterial genome structures.
\newblock {\em Computer Graphics Forum}, 38, 2019.

\bibitem{wang1961proving}
Hao Wang.
\newblock Proving theorems by pattern recognition{-II}.
\newblock {\em Bell Labs Technical Journal}, 40(1):1--41, 1961.

\bibitem{fu2005texture}
Chi-Wing Fu and Man-Kang Leung.
\newblock Texture tiling on arbitrary topological surfaces using {Wang} tiles.
\newblock In {\em Rendering Techniques}, pages 99--104, 2005.

\bibitem{Li-Yi-2004-Tile-Based-Texture-Mapping-on-Graphics-Hardware}
Li-Yi Wei.
\newblock Tile-based texture mapping on graphics hardware.
\newblock In {\em ACM SIGGRAPH 2004 Sketches}, SIGGRAPH '04, page~67, New York,
  NY, USA, 2004. Association for Computing Machinery.

\bibitem{Culik-1995-An-Aperiodic-Set-of-Wang-Cubes}
Karel Cul{\'i}k and Jarkko Kari.
\newblock An aperiodic set of wang cubes.
\newblock {\em J. Univers. Comput. Sci.}, 1:675--686, 1995.

\bibitem{Doskar-2020-Level-set-Based-Design-of-Wang-Tiles-for-Modelling-Complex-Microstructures}
Martin Doškář, Jan Zeman, Daniel Rypl, and Jan Novák.
\newblock {Level-set Based Design of Wang Tiles for Modelling Complex
  Microstructures}.
\newblock {\em Computer-Aided Design}, 123:102827, 2020.

\bibitem{fleischer1995cellular}
Kurt~W Fleischer, David~H Laidlaw, Bena~L Currin, and Alan~H Barr.
\newblock Cellular texture generation.
\newblock In {\em Proceedings of the 22nd annual conference on Computer
  graphics and interactive techniques}, pages 239--248, 1995.

\bibitem{VMD}
William Humphrey, Andrew Dalke, and Klaus Schulten.
\newblock {VMD: Visual molecular dynamics}.
\newblock {\em Journal of Molecular Graphics}, 14(1):33--38, 1996.

\bibitem{PyMOL}
{Schr\"odinger, LLC}.
\newblock {The {PyMOL} Molecular Graphics System, Version~1.8}.
\newblock November 2015.

\bibitem{Knoll-2013-Ray-Tracing-and-Volume-Rendering-Large-Molecular-Data-on-Multi-Core-and-Many-Core-Architectures}
Aaron Knoll, Ingo Wald, Paul~A. Navr\'{a}til, Michael~E. Papka, and Kelly~P.
  Gaither.
\newblock Ray tracing and volume rendering large molecular data on multi-core
  and many-core architectures.
\newblock In {\em Proceedings of the 8th International Workshop on Ultrascale
  Visualization}, UltraVis '13, New York, NY, USA, 2013. Association for
  Computing Machinery.

\bibitem{Amira}
{Amira}.
\newblock [Online].

\bibitem{Waltemate-2014-Membrane-Mapping}
Thomas Waltemate, Björn Sommer, and Mario Botsch.
\newblock {Membrane Mapping: Combining Mesoscopic and Molecular Cell
  Visualization}.
\newblock In {\em Eurographics Workshop on Visual Computing for Biology and
  Medicine}, 2014.

\bibitem{Sehnal21}
David Sehnal, Sebastian Bittrich, Mandar Deshpande, Radka Svobodová, Karel
  Berka, Václav Bazgier, Sameer Velankar, Stephen~K Burley, Jaroslav Koča,
  and Alexander~S Rose.
\newblock {Mol* Viewer: modern web app for 3D visualization and analysis of
  large biomolecular structures}.
\newblock {\em Nucleic Acids Research}, 49(W1):W431--W437, 05 2021.

\bibitem{Grottel-2015-MegaMol}
Sebastian Grottel, Michael Krone, Christoph Müller, Guido Reina, and Thomas
  Ertl.
\newblock {MegaMol—A Prototyping Framework for Particle-Based Visualization}.
\newblock {\em IEEE Transactions on Visualization and Computer Graphics},
  21(2):201--214, 2015.

\bibitem{OSMegamol}
Sebastian Grottel, Michael Krone, Katrin Scharnowski, and Thomas Ertl.
\newblock Object-space ambient occlusion for molecular dynamics.
\newblock In {\em 2012 IEEE Pacific Visualization Symposium}, pages 209--216,
  2012.

\bibitem{Ibrahim-2021-ProbabilisticParticleOcclusionCulling}
Mohamed Ibrahim, Peter Rautek, Guido Reina, Marco Agus, and Markus Hadwiger.
\newblock {Probabilistic Occlusion Culling using Confidence Maps for
  High-Quality Rendering of Large Particle Data}.
\newblock {\em IEEE Transactions on Visualization and Computer Graphics
  (Proceedings IEEE VIS 2021)}, 28(1):573--582, 2022.

\bibitem{Lindow-2012-Interactive-Rendering-Of-Materials}
N.~Lindow, D.~Baum, and H.-C. Hege.
\newblock {Interactive Rendering of Materials and Biological Structures on
  Atomic and Nanoscopic Scale}.
\newblock {\em Comput. Graph. Forum}, 31(3pt4):1325–1334, jun 2012.

\bibitem{Falk-2013-Atomistic-Visualization-of-Mesoscopic}
Martin Falk, Michael Krone, and Thomas Ertl.
\newblock {Atomistic Visualization of Mesoscopic Whole‐Cell Simulations Using
  Ray‐Casted Instancing}.
\newblock {\em Computer Graphics Forum}, 32, 2013.

\bibitem{LeMuzic-2014-Illustrative-Visualization-of-Molecular-Reactions-using-Omniscient-Intelligence}
Mathieu~Le Muzic, Julius Parulek, Anne-Kristin Stavrum, and Ivan Viola.
\newblock {Illustrative Visualization of Molecular Reactions using Omniscient
  Intelligence and Passive Agents }.
\newblock {\em Computer Graphics Forum}, 33(3):141--150, June 2014.
\newblock Article first published online: 12 JUL 2014.

\bibitem{LeMuzic-2015-cellVIEW}
Mathieu~Le Muzic, Ludovic Autin, J{\'u}lius Parulek, and Ivan Viola.
\newblock {cellVIEW: a Tool for Illustrative and Multi-Scale Rendering of Large
  Biomolecular Datasets}.
\newblock {\em Eurographics Workshop on Visual Computing for Biomedicine},
  2015:61--70, 2015.

\bibitem{Tarini06}
Marco Tarini, Paolo Cignoni, and Claudio Montani.
\newblock Ambient occlusion and edge cueing for enhancing real time molecular
  visualization.
\newblock {\em IEEE Transactions on Visualization and Computer Graphics},
  12(5):1237--1244, 2006.

\bibitem{Michel20}
{\'{E}}lie Michel and Tamy Boubekeur.
\newblock Real time multiscale rendering of dense dynamic stackings.
\newblock {\em Computer Graphics Forum}, 39(7):169--179, 2020.

\bibitem{OSPRay}
I~Wald, GP~Johnson, J~Amstutz, C~Brownlee, A~Knoll, J~Jeffers, J~Günther, and
  P~Navratil.
\newblock Ospray - a cpu ray tracing framework for scientific visualization.
\newblock {\em IEEE Transactions on Visualization and Computer Graphics},
  23(1):931--940, 2017.

\bibitem{Embree}
Ingo Wald, Sven Woop, Carsten Benthin, Gregory~S. Johnson, and Manfred Ernst.
\newblock Embree: A kernel framework for efficient cpu ray tracing.
\newblock {\em ACM Trans. Graph.}, 33(4), jul 2014.

\bibitem{RTXpkd}
Patrick Gralka, Ingo Wald, Sergej Geringer, Guido Reina, and Thomas Ertl.
\newblock Spatial partitioning strategies for memory-efficient ray tracing of
  particles.
\newblock In {\em 2020 IEEE 10th Symposium on Large Data Analysis and
  Visualization (LDAV)}, pages 42--52, 2020.

\bibitem{Molnar-1994-A-sorting-classification-of-parallel-rendering}
S.~Molnar, M.~Cox, D.~Ellsworth, and H.~Fuchs.
\newblock A sorting classification of parallel rendering.
\newblock {\em IEEE Computer Graphics and Applications}, 14(4):23--32, 1994.

\bibitem{Molnar-1992-PixelFlow-High-Speed-Rendering-Using-Image-Composition}
Steven Molnar, John Eyles, and John Poulton.
\newblock {PixelFlow: High-Speed Rendering Using Image Composition}.
\newblock In {\em Proceedings of the 19th Annual Conference on Computer
  Graphics and Interactive Techniques}, SIGGRAPH '92, page 231–240, New York,
  NY, USA, 1992. Association for Computing Machinery.

\bibitem{Eilemann-2019-Parallel-Rendering-and-Large-Data-Visualization}
Stefan Eilemann.
\newblock {Parallel Rendering and Large Data Visualization}.
\newblock {\em CoRR}, abs/1902.08755, 2019.

\bibitem{Zellmann-2020-Finding-Efficient-Spatial-Distributions-for-Massively-Instanced-3-d-Models}
Stefan Zellmann, Nate Morrical, Ingo Wald, and Valerio Pascucci.
\newblock {Finding Efficient Spatial Distributions for Massively Instanced 3-d
  Models}.
\newblock In Steffen Frey, Jian Huang, and Filip Sadlo, editors, {\em
  Eurographics Symposium on Parallel Graphics and Visualization}. The
  Eurographics Association, 2020.

\bibitem{Wang1961}
Hao Wang.
\newblock Proving theorems by pattern recognition — ii.
\newblock {\em The Bell System Technical Journal}, 40(1):1--41, 1961.

\bibitem{LipidWrapper}
Jacob~D. Durrant and Rommie~E. Amaro.
\newblock Lipidwrapper: An algorithm for generating large-scale membrane models
  of arbitrary geometry.
\newblock {\em PLOS Computational Biology}, 10(7):1--11, 07 2014.

\bibitem{WangTilesforImageandTextureGeneration}
Michael~F. Cohen, Jonathan Shade, Stefan Hiller, and Oliver Deussen.
\newblock Wang tiles for image and texture generation.
\newblock {\em ACM Trans. Graph.}, 22(3):287–294, jul 2003.

\bibitem{An-aperiodic-set-of-11-Wang-tiles}
Emmanuel Jeandel and Micha{\"{e}}l Rao.
\newblock An aperiodic set of 11 wang tiles.
\newblock {\em CoRR}, abs/1506.06492, 2015.

\bibitem{Wald-2001-State-of-the-Art-in-Interactive-RayTracing}
Ingo Wald and Philipp Slusallek.
\newblock {State of the Art in Interactive Ray Tracing}.
\newblock In {\em Eurographics 2001 - STARs}. Eurographics Association, 2001.

\bibitem{Wald-2001-Interactive-Distributed-RayTracing-Highly-Complex-Models}
Ingo Wald, Philipp Slusallek, and Carsten Benthin.
\newblock {Interactive Distributed Ray Tracing of Highly Complex Models}.
\newblock In {\em Proceedings of the 12th Eurographics Conference on
  Rendering}, EGWR'01, page 277–288, Goslar, DEU, 2001. Eurographics
  Association.

\bibitem{Wald-2002-OpenRT}
Ingo Wald.
\newblock {A Flexible and Scalable Rendering Engine for Interactive 3D
  Graphics}.
\newblock 2002.

\bibitem{RayTracingGems}
Eric Haines and Tomas Akenine-M\"oller, editors.
\newblock {\em Ray Tracing Gems}.
\newblock Apress, 2019.
\newblock \url{http://raytracinggems.com}.

\bibitem{RayQueries}
{khronos Group}.
\newblock {Ray Tracing In Vulkan}, 2020.
\newblock \url{https://www.khronos.org/blog/ray-tracing-in-vulkan}.

\bibitem{aliaga-1999-Automatic-image-placement-to-provide-guaranteed-frame-rate}
Daniel~G Aliaga and Anselmo Lastra.
\newblock {Automatic image placement to provide a guaranteed frame rate}.
\newblock In {\em Proceedings of the 26th annual conference on Computer
  graphics and interactive techniques}, pages 307--316, 1999.

\bibitem{Aliaga-1999-MMR}
Daniel~G. Aliaga, Jonathan~D. Cohen, Andrew~T. Wilson, Eric Baker, Hansong
  Zhang, Carl Erikson, Kenneth~E. Hoff, Thomas~C. Hudson, Wolfgang
  Stuerzlinger, Rui Bastos, Mary~C. Whitton, Frederick~P. Brooks, and Dinesh
  Manocha.
\newblock {MMR: an interactive massive model rendering system using geometric
  and image-based acceleration}.
\newblock In {\em SI3D}, 1999.

\bibitem{schaufler-1996-Three-Dimensional-Image-Cache}
Gernot Schaufler and Wolfgang Stürzlinger.
\newblock Three dimensional image cache for virtual reality.
\newblock {\em Computer Graphics Forum}, 15(3):227--235, Aug 1996.
\newblock Eurographics '96.

\bibitem{VULKAN}
Khronos~Group Inc.
\newblock {Ray Tracing in Vulkan}, 2020.

\bibitem{nvpro-samples}
NVIDIA.
\newblock {Nvpro-core: NVIDIA DesignWorks Samples}, Accessed Feb. 23, 2021.
\newblock \url{https://github.com/nvpro-samples}.

\bibitem{Mindek-2018-marion}
Peter Mindek, David Kouřil, Johannes Sorger, Daniel Toloudis, Blair Lyons,
  Graham Johnson, M.~Eduard Gröller, and Ivan Viola.
\newblock Visualization multi-pipeline for communicating biology.
\newblock {\em IEEE Transactions on Visualization and Computer Graphics},
  24(1):883--892, 2018.

\bibitem{NVIDIA-2021-Ampere-GPU-Architecture-whitepaper}
NVIDIA.
\newblock {Ampere GA102 GPU Architecture whitepaper}, 2021.

\bibitem{NVIDIA-Ambient-Occlusion}
{NVIDIA}.
\newblock {G-Buffer and Ambient Occlusion - Tutorial}.
\newblock
  \url{https://github.com/nvpro-samples/vk_raytracing_tutorial_KHR/tree/master/ray_tracing_ao}.

\bibitem{Dietrich-2007-Massive-Model-Rendering-Techniques}
Andreas Dietrich, Enrico Gobbetti, and Sung-Eui Yoon.
\newblock Massive-model rendering techniques: A tutorial.
\newblock {\em IEEE Computer Graphics and Applications}, 27(6):20--34, 2007.

\bibitem{Avizo}
Peter Westenberger.
\newblock Avizo—three-dimensional visualization framework.
\newblock In {\em Geoinformatics 2008—Data to Knowledge}, pages 13--14. USGS,
  2008.

\bibitem{paraview}
Utkarsh Ayachit.
\newblock {\em The paraview guide: a parallel visualization application}.
\newblock Kitware, Inc., 2015.

\bibitem{OSPRayStudio2021}
Isha Sharma, Dave DeMarle, Alok Hota, Bruce Cherniak, and Johannes Günther.
\newblock {OSPRay Studio: Enabling Multi-Workflow Visualizations with OSPRay}.
\newblock In Christina Gillmann, Michael Krone, Guido Reina, and Thomas
  Wischgoll, editors, {\em VisGap - The Gap between Visualization Research and
  Visualization Software}. The Eurographics Association, 2021.

\bibitem{Thonat-2021-Tessellation-Free-Displacement-Mapping-for-Ray-Tracing}
Theo Thonat, Francois Beaune, Xin Sun, Nathan Carr, and Tamy Boubekeur.
\newblock {Tessellation-Free Displacement Mapping for Ray Tracing}.
\newblock {\em ACM Trans. Graph.}, 40(6), dec 2021.

\end{thebibliography}

\end{document}